\documentclass[preprints,article,accept,pdftex,oneauthor]{Definitions/mdpi}
\firstpage{1} 
\pubvolume{16}
\issuenum{11}
\articlenumber{1520}
\pubyear{2024}
\copyrightyear{2024}
\externaleditor{\textls[-25]{Academic Editor: Manuel Gadella}}
\datereceived{24 October 2024 } 
\daterevised{8 November 2024} % Comment out if no revised date
\dateaccepted{ 8 November 2024} 
\datepublished{13 November 2024} 
\hreflink{https://doi.org/\\10.3390/sym16111520} % If needed use \linebreak
\pdfoutput=1 % Uncommented for upload to arXiv.org
\everymath{\displaystyle}

\newcommand{\dd}{\mathrm{d}}
\newcommand{\vd}[1]{\dd{#1}\,}
\newcommand{\ud}{\,\dd}
\newcommand{\nbr}[1]{$#1$\nobreakdash-\hspace{0pt}}
\newcommand{\braket}[1]{\langle{#1}\rangle}

\newcommand{\braketket}[1]{\langle{#1}\rangle\!\rangle}
\newcommand{\ket}[1]{\lvert{#1}\rangle}
\newcommand{\bra}[1]{\langle{#1}\rvert}
\newcommand{\ketket}[1]{\lvert{#1}\rangle\!\rangle}
\newcommand{\brabra}[1]{\langle\!\langle{#1}\rvert}
\newcommand{\sdot}{\,\cdot\,}
\providecommand{\abs}[1]{\lvert#1\rvert}

\providecommand{\norm}[1]{\lVert#1\rVert}

\DeclareMathOperator{\Tr}{Tr}
\DeclareMathOperator{\re}{Re}
\DeclareMathOperator{\im}{Im}
\DeclareMathOperator{\sinc}{sinc}
\DeclareMathOperator{\artanh}{ar\,tanh}
\DeclareMathOperator{\Prob}{Prob}

\Title{Time Evolution in Quantum Mechanics with a Minimal Time~Scale}
\TitleCitation{Time Evolution in Quantum Mechanics with a Minimal Time Scale}

\Author{Ziemowit Doma\'nski \orcidA{}}
\AuthorNames{Ziemowit Doma\'nski}
\AuthorCitation{Doma\'nski, Z.}

\address[1]{Institute of Mathematics, Faculty of Control, Robotics and
Electrical Engineering, Pozna\'n University of Technology, Piotrowo 3A,
60-965 Pozna\'n, Poland; ziemowit.domanski@put.poznan.pl}

\abstract{The existence of a minimum measurable length scale was suggested by
various theories of quantum gravity, string theory and black hole physics.
Motivated by this, we examine a quantum theory exhibiting a minimum measurable
time scale. We use the Page--Wootters formalism to describe time evolution of a
quantum system with the modified commutation relations between the time and
frequency operator. Such modification leads to a minimal uncertainty in the
measurement of time. This causes breaking of the time-translation symmetry and
results in a modified version of the Schr\"odinger equation. A minimal time
scale also allows us to introduce a discrete Schr\"odinger equation describing
time evolution on a lattice. We show that both descriptions of time evolution
are equivalent. We demonstrate the developed theory on a couple simple quantum
systems.}

\keyword{\textls[-15]{quantum mechanics; minimal time scale;
Page--Wootters formalism; generalized  uncertainty principle}}

\begin{document}

\section{Introduction}
\label{sec:1}
Various theories of quantum gravity (such as string theory) predict the
existence of a minimum measurable length scale, usually on the order of the
Planck length \cite{Amati:1989,Garay:1995,Hossenfelder:2003,Ali:2009}. Black
hole physics also suggests the existence of a fundamental limit with which we
can measure distances \cite{Maggiore:1993,Scardigli:1999,Hossenfelder:2013}.
This is because the energy needed to probe spacetime below the Planck length
scale exceeds the energy needed to produce a black hole in that region of
spacetime. On the other hand, Doubly Special Relativity theories predict maximum
observable momenta \cite{Magueijo:2002,Cortes:2005}. All of these theories give
rise to the modified commutation relations between position coordinates and
momenta, which in turn give rise to the Generalized Uncertainty Principle.

Motivated by these results, we examine a quantum theory exhibiting a minimum
measurable time scale. Our starting point is the Page--Wootters formalism of
time evolution in quantum mechanics. It is based on the idea that time evolution
arises as a result of correlations between the ``clock'' and the rest of the
system. Next, we introduce the minimal time scale into the theory by modifying
the commutation relations between the time operator $\hat{T}$ and the operator
conjugate to it (the frequency operator $\hat{\Omega}$). Such a modification
causes the time observable to not be represented by a self-adjoint operator
but only by a symmetric operator. This is expected as it is no longer possible
to consider states of the system at a particular instance of time. As a result
we use states of the ``clock'' system maximally localized around instances of
time to construct continuous time representation of the system and we arrive at
a modified Schr\"odinger equation describing time evolution of the system.
A minimal time scale also allows us to introduce a discrete time representation
of the system and a corresponding discrete Schr\"odinger equation describing
time evolution on a lattice. Interestingly, the received discrete and continuous
time representations are equivalent with each other and describe the same time
evolution of the system.

According to the received Schr\"odinger Equations \eqref{eq:15} and
\eqref{eq:16}, time evolution is governed by an effective Hamiltonian different
than the Hamiltonian of the system. The reason for this is that in the case
without a minimal time scale, the isolated system possesses time-translation
symmetry. The Hamiltonian is the generator of this symmetry. However, when there
exists a minimum measurable time scale, the time-translation symmetry breaks
down and the Hamiltonian no longer generates translations in time.

Other approaches to time evolution with a minimal time scale in quantum
mechanics can be found in the literature. Worth noting is the paper
\cite{Faizal:2016}, where the authors deform the Heisenberg algebra of spacetime
variables receiving a modified Schr\"odinger equation
\begin{equation}
i\hbar\frac{\partial}{\partial t} \ket{\psi}
+ \hbar^2 \alpha \frac{\partial^2}{\partial t^2} \ket{\psi} = \hat{H}\ket{\psi},
\end{equation}
where $\alpha$ is the deformation parameter. Interesting results were also
received in \cite{Faizal:2017}, where a deformation of the Wheeler--DeWitt
equation led to a discretization of time. The paper~\cite{Caldirola:1956} by
P.~Caldirola should also be mentioned, where the author postulates the existence
of a universal interval $\tau_0$ of proper time. The existence of such a
universal interval causes the reaction of a particle to the applied external
force to not be continuous. As a result, positions of the particle along its
world line are discretized.

Breaking of time-translation symmetry is a basic requirement for the
creation of time crystals \cite{Wilczek:2012,Sacha:2015,Yao:2017}. Therefore,
minimal time scales can lead to a behavior of the system undergoing time
evolution similar to that of a time crystal \cite{Faizal:2016,Faizal:2017}.

This paper is structured in the following way. In Section~\ref{sec:2}, we review
the Page--Wootters formalism on which the developed theory will be based. In
Section~\ref{sec:3}, we introduce a minimal time scale into the theory by
modifying the commutation relations between the time and frequency operators.
Section~\ref{sec:4} contains the analysis of the time evolution of a couple
simple quantum systems when a minimal time scale is present. The systems of one,
two and three spins-$1/2$ in an external magnetic field are investigated, as
well as a free particle and a harmonic oscillator. The conclusions and a
discussion of the received results is given in Section~\ref{sec:5}.

\section{Page--Wootters Formalism}
\label{sec:2}
In a standard description of quantum mechanics, time is treated as a background
structure with respect to which quantization is performed. In the Schr\"odinger
equation, time is treated as a classical parameter. Physically, it represents
the time shown by a ``classical'' clock in the laboratory. Moreover, in
relativistic mechanics, standard quantization techniques are imposing the
canonical commutation relations on constant-time hypersurfaces. Furthermore,
the Wheeler--DeWitt equation \cite{DeWitt:1967} says
\begin{equation}
\hat{H}\ket{\psi} = 0,
\label{eq:27}
\end{equation}
where $\hat{H}$ is the Hamiltonian constraint in quantized general relativity
and $\ket{\psi}$ stands for the wave function of the universe. The universe as
a whole is static and does not evolve.

Therefore, it is important to give time a fully quantum description so that
time becomes a quantum degree of freedom. In the following section, we will
review the Page--Wootters formalism \cite{Page:1983,Gambini:2009,%
Giovannetti:2015,Hohn:2021}, which accomplishes this and which will be
particularly suited for the considerations presented in this paper. Other
proposals of quantum descriptions of time can be found in the literature
\cite{Salecker:1958,Zeh:1986,Rovelli:1991,Kuchar:2011,Aharonov:2014,Gozdz:2023,%
Gozdz:2024}.

In the Page--Wootters formalism, one assigns to the time degree of freedom a
Hilbert space $\mathcal{H}_T$. This Hilbert space describes a ``clock'', which
will measure the flow of time. The system undergoing the time evolution will be
described by the Hilbert space $\mathcal{H}_S$. The joint Hilbert space of the
``clock'' and the system is
\begin{equation}
\mathcal{H} = \mathcal{H}_T \otimes \mathcal{H}_S.
\end{equation}

We will assume that $\mathcal{H}_T$ is isomorphic to the Hilbert space of a
particle on a line, i.e.,\ the space $L^2(\mathbb{R},\dd{x})$ of complex-valued
functions defined on $\mathbb{R}$ and square integrable with respect to the
Lebesgue measure. It is worth noting that other choices are also
possible~\mbox{\cite{Wootters:1984,Moreva:2014}}. On $\mathcal{H}_T$, we
introduce the time operator $\hat{T}$ corresponding to the measurements of time
and the frequency operator $\hat{\Omega}$ conjugate to $\hat{T}$:
\begin{equation}
[\hat{T},\hat{\Omega}] = i\hat{1}.
\end{equation}

When $\mathcal{H}_T \cong L^2(\mathbb{R},\dd{x})$, then the operators $\hat{T}$
and $\hat{\Omega}$ will be represented as the operators of multiplication and
differentiation:
\begin{equation}
\hat{T}\psi(t) = t\psi(t), \quad
\hat{\Omega}\psi(t) = -i\partial_t \psi(t).
\label{eq:29}
\end{equation}

On $\mathcal{H}$, we introduce the constraint operator of the model
\begin{equation}
\hat{J} = \hbar\hat{\Omega} \otimes \hat{1}_S + \hat{1}_T \otimes \hat{H}_S,
\label{eq:30}
\end{equation}
with $\hat{H}_S$ the system Hamiltonian. Such constraint describes the simplest
case of the ``clock'' non-interacting with the system and where the system
Hamiltonian is time independent. It is, however, possible to consider more
general cases. Since $\hat{\Omega}$ and $\hat{H}_S$ are self-adjoint operators,
the constraint operator $\hat{J}$ will be also self-adjoint. If we treat
$\hat{J}$ as a total Hamiltonian, then in accordance with the Wheeler--DeWitt
Equation \eqref{eq:27} we define physical states of the model as vectors
$\ketket{\Psi}$ satisfying
\begin{equation}
\hat{J}\ketket{\Psi} = 0.
\label{eq:28}
\end{equation}

We will use the double-ket notation to denote states $\ketket{\Psi}$ from the
total Hilbert space $\mathcal{H}$. Formula \eqref{eq:28} defines physical states
as (generalized) eigenvectors of the operator $\hat{J}$ associated with the null
eigenvalue. The physical states $\ketket{\Psi}$ provide a complete description
of the temporal evolution of the system $S$ by representing it in terms of
correlations (entanglement) between the latter and the ``clock'' system.

The conventional state $\ket{\psi(t)}_S$ of the system $S$ at time $t$ can be
obtained via projection of $\ketket{\Psi}$ with a generalized eigenstate
$\ket{t}_T$ of the time operator $\hat{T}$:
\begin{equation}
\ket{\psi(t)}_S = {}_T\braketket{t|\Psi},
\end{equation}
where $\ket{t}_T$ is the generalized eigenvector of the operator $\hat{T}$
associated with the eigenvalue~$t$:
\begin{equation}
\hat{T}\ket{t}_T = t\ket{t}_T.
\end{equation}

By writing \eqref{eq:28} in the time representation in $\mathcal{H}_T$ and with
the help of \eqref{eq:29}:
\begin{equation}
{}_T\bra{t}\hat{J}\ketket{\Psi} = 0 \quad\iff\quad
i\hbar\frac{\partial}{\partial t}\ket{\psi(t)}_S = \hat{H}_S \ket{\psi(t)}_S
\end{equation}
we verify that the vectors $\ket{\psi(t)}_S$ obey the Schr\"odinger equation.

By projecting $\ketket{\Psi}$ on the generalized eigenstates $\ket{\omega}_T$ of
$\hat{\Omega}$:
\begin{equation}
\ket{\psi(\omega)}_S = {}_T\braketket{\omega|\Psi},
\end{equation}
where $\ket{\omega}_T$ is the generalized eigenvector of the operator
$\hat{\Omega}$ associated with the eigenvalue~$\omega$:
\begin{equation}
\hat{\Omega}\ket{\omega}_T = \omega\ket{\omega}_T,
\end{equation}
we obtain the eigenvector equation of the operator $\hat{H}_S$:
\begin{equation}
{}_T\bra{\omega}\hat{J}\ketket{\Psi} = 0 \quad\iff\quad
\hat{H}_S\ket{\psi(\omega)}_S = -\hbar\omega\ket{\psi(\omega)}_S.
\end{equation}

Specifically, for $\omega$ such that $-\hbar\omega$ equals an element of the
spectrum of $\hat{H}_S$, then $\ket{\psi(\omega)}_S$ is an eigenvector of
$\hat{H}_S$ at that eigenvalue; otherwise, $\ket{\psi(\omega)}_S = 0$.

When the total system is in a state $\ketket{\Psi}$, the probability that the
measurement of an observable $\hat{A}_S$ (a~self-adjoint operator defined on the
Hilbert space $\mathcal{H}_S$) will give a value $a$ from the spectrum of
$\hat{A}_S$ given that the ``clock'' reads the time $t$ is postulated as the
conditional probability
\begin{equation}
\Prob(\text{$a$ when $t$}) = \frac{\Prob(\text{$a$ and $t$})}{\Prob(t)}
= \frac{\brabra{\Psi}(\ket{t}\bra{t}_T \otimes \ket{a}\bra{a}_S)\ketket{\Psi}}{\brabra{\Psi}(\ket{t}\bra{t}_T \otimes \hat{1}_S)\ketket{\Psi}}
= \frac{\abs{{}_S\braket{a|\psi(t)}_S}^2}{{}_S\braket{\psi(t)|\psi(t)}_S},
\end{equation}
where $\ket{a}_S$ is an eigenvector of the operator $\hat{A}_S$ associated with
an eigenvalue $a$. We can see that the above postulate reproduces the standard
Born rule of computing the probabilities of measurements.

\begin{Remark}
The original formulation of the formalism by Page and Wootters was criticized as
it seemed that it is not reproducing the correct formulas when calculating
the probability of measuring an observable $\hat{B}_S$ at time $t_2$ when one
finds the system at an eigenstate of an observable $\hat{A}_S$ at time $t_1$.
In other words we are not receiving the correct propagators of subsequent
measurements of observables $\hat{A}_S$ and $\hat{B}_S$. For the possible
resolution of this problem, see \cite{Gambini:2009,Giovannetti:2015,Hohn:2021}.
\end{Remark}

\begin{Remark}
The Page--Wootters formalism can be extended to incorporate a time-dependent
Hamiltonians by replacing the constraint operator \eqref{eq:30} with
\begin{equation}
\hat{J} = \hbar\hat{\Omega} \otimes \hat{1}_S + \hat{H}_S(\hat{T}),
\label{eq:33}
\end{equation}
where $\hat{H}_S(\hat{T})$ is an operator on $\mathcal{H}$ that explicitly
depends on the time operator $\hat{T}$.
\end{Remark}

\begin{Remark}
The generalized eigenvectors of operators appearing in the Page--Wootters
formalism are formally defined using rigged Hilbert spaces. A rigged Hilbert
space consists of a Hilbert space $\mathcal{H}$, together with a dense subspace
$\Phi$, such that $\Phi$ is given a topological vector space structure with a
finer topology than the one inherited from $\mathcal{H}$; i.e.,\ the inclusion
map $\imath \colon \Phi \to \mathcal{H}$ is continuous. By $\Phi^*$, we will
denote the dual space to $\Phi$, i.e.,\ the space of continuous linear
functionals $\Phi \to \mathbb{C}$. Similarly, by $\mathcal{H}^*$ we will denote
the dual space to $\mathcal{H}$. The adjoint map $\imath^* \colon \mathcal{H}^*
\to \Phi^*$ to $\imath$ defined by
\begin{equation}
\braket{\imath^*(\varphi),\psi} = \braket{\varphi,\imath(\psi)}, \quad
\varphi \in \mathcal{H}^*, \psi \in \Phi
\label{eq:32}
\end{equation}
is injective by the density of $\Phi$ in $\mathcal{H}$. Therefore, $\imath^*$
provides us with an inclusion of $\mathcal{H}^*$ in $\Phi^*$. By identifying
$\mathcal{H}^*$ with $\mathcal{H}$ in accordance to the Riesz representation
theorem we obtain the inclusion of $\mathcal{H}$ in $\Phi^*$, i.e.,
\begin{equation}
\Phi \subset \mathcal{H} \subset \Phi^*.
\end{equation}

By \eqref{eq:32}, the duality pairing $\braket{\sdot,\sdot}$ between $\Phi$ and
$\Phi^*$ is compatible with the inner product $(\sdot,\sdot)$ on $\mathcal{H}$:
\begin{equation}
\braket{\varphi,\psi} = (\varphi,\psi), \quad
\varphi \in \mathcal{H} \subset \Phi^*, \psi \in \Phi \subset \mathcal{H}.
\end{equation}

The space $\Phi$ plays the role of the space of test functions and $\Phi^*$ is
the space of distributions (generalized vectors).

As an example let $\mathcal{H} = L^2(\mathbb{R},\dd{x})$ and $\Phi$ be the
Schwartz space of rapidly decreasing functions. Then, $\Phi^*$ is the space of
tempered distributions. The operators $\hat{T}$ and $\hat{\Omega}$ given by
\eqref{eq:29} restricted to $\Phi$ also take values in $\Phi$. They are also
continuous with respect to the topology in $\Phi$. Therefore, they can be
extended to operators on $\Phi^*$ and we obtain the following eigenvector
equations for these~operators
\begin{equation}
\hat{T}\delta(t - t_0) = t_0\delta(t - t_0), \quad
\hat{\Omega}e^{i\omega t} = \omega e^{i\omega t}.
\end{equation}

Both $\delta(t - t_0)$ and $e^{i\omega t}$ are generalized vectors as they are
not elements in $L^2(\mathbb{R},\dd{x})$.
\end{Remark}

\section{Incorporation of a Minimal Time Scale}
\label{sec:3}
A minimal time scale can be incorporated into the formalism by modifying the
commutation relations for the operators of time and frequency, $\hat{T}$ and
$\hat{\Omega}$:
\begin{equation}
[\hat{T},\hat{\Omega}] = i\left(\hat{1} + \kappa\hat{\Omega}^2\right).
\label{eq:1}
\end{equation}

Here $\kappa$ is a positive constant describing the smallest possible resolution
with which we can measure time. For $\kappa = 0$, we recover the standard
commutation relations for $\hat{T}$ and $\hat{\Omega}$. The commutation
relations of the form \eqref{eq:1} for the operators of position and momentum
were considered in \cite{Kempf:1995} from the point of view of a minimal length
scale in quantum mechanics. In the following subsection, we will review some of
the results from this paper which will be crucial for our later considerations.

\subsection{Minimal Time Scale Uncertainty Relation}
\label{subsec:3.1}
The commutation relations \eqref{eq:1} imply the following generalized
uncertainty principle for the uncertainties $\Delta t$ and $\Delta\omega$ of
the measurements of time and frequency:
\begin{equation}
\Delta t \Delta \omega \geq \frac{1}{2}\left(1 + \kappa (\Delta \omega)^2
    + \kappa \braket{\hat{\Omega}}^2\right).
\label{eq:2}
\end{equation}

\begin{figure}[H]
\includegraphics{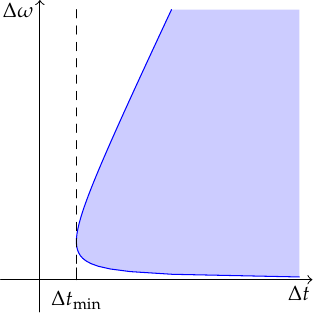}
%\begin{asy}
%usepackage("mathpazo");
%import graph;
%size(150,IgnoreAspect);
%defaultpen(fontsize(10pt));

%real a = 1;
%pair f(real t) { return ((1 + t*t + a*a)/(2*t),t); }

%fill(graph(f, 1/10, 10)--(f(1/10).x,f(10).y)--cycle,rgb(0.8,0.8,1));
%draw(graph(f, 1/10, 10),blue);
%draw((sqrt(1 + a*a),0)--(sqrt(1 + a*a),f(10).y),dashed);
%xaxis("$\Delta t$",Arrow(TeXHead));
%yaxis("$\Delta \omega$",Arrow(TeXHead),autorotate=false);
%xtick("$\Delta t_{\mathrm{min}}$",sqrt(1 + a*a));
%\end{asy}
\caption{Plot of the inequality in \eqref{eq:2} showing that there exists a
smallest possible value for the uncertainty $\Delta t$.}
\label{fig:1}
\end{figure}

From \eqref{eq:2}, it is not difficult to see that there exists a smallest
possible value for the uncertainty $\Delta t$ (see Figure~\ref{fig:1}):
\begin{equation}
\Delta t_{\mathrm{min}} = \sqrt{\kappa\bigl(1 + \kappa\braket{\hat{\Omega}}^2\bigr)}.
\end{equation}

Every physical state cannot have the uncertainty in time smaller than this
value. For states for which the expectation value of the frequency operator
$\hat{\Omega}$ is equal zero, we receive the absolutely smallest uncertainty in
time equaling
\begin{equation}
\Delta t_0 = \sqrt{\kappa}.
\end{equation}

It should be noted that there are states in the Hilbert space $\mathcal{H}_T$,
like, for example, the eigenstates of the time operator $\hat{T}$, for which the
uncertainty in time is smaller than $\Delta t_{\mathrm{min}}$. Such states are
not considered physical.

The modification \eqref{eq:1} of the commutation relations for $\hat{T}$ and
$\hat{\Omega}$ is the simplest one leading to a minimal uncertainty in time.
Other modifications are possible; however, we will not be dealing with such
cases in the current work.

The introduction of a minimal uncertainty in time will cause the time operator
$\hat{T}$ to not be essentially self-adjoint but only symmetric. A consequence
of this is that eigenvectors of $\hat{T}$ are not physical states. It is no
longer possible to consider states of the system $S$ at particular instances of
time. Therefore, we do not have at a disposal the representation of the Hilbert
space $\mathcal{H}_T$ given by projecting states $\ket{\psi}_T$ onto
eigenvectors $\ket{t}_T$ of $\hat{T}$. However, the frequency operator
$\hat{\Omega}$ is still self-adjoint, and therefore we can use it to construct a
frequency representation of the Hilbert space $\mathcal{H}_T$. The Hilbert space
$\mathcal{H}_T$ is represented as a space $L^2(\mathbb{R},\dd{\mu})$ of
functions defined on $\mathbb{R}$ and square integrable with respect to the
measure $\dd{\mu}(\omega) = \frac{\dd\omega}{1 + \kappa\omega^2}$. Such
functions can be formally constructed by projecting states $\ket{\psi}_T$ onto
eigenvectors $\ket{\omega}_T$ of $\hat{\Omega}$:
\begin{equation}
\psi(\omega) = {}_T\braket{\omega|\psi}_T.
\end{equation}

The scalar product in the Hilbert space $L^2(\mathbb{R},\dd{\mu})$ is given by
the formula
\begin{equation}
{}_T\braket{\varphi|\psi}_T = \int_{-\infty}^{+\infty}
    \frac{\dd\omega}{1 + \kappa\omega^2} \overline{\varphi(\omega)}\psi(\omega),
\end{equation}
and the frequency and time operators $\hat{\Omega}$, $\hat{T}$ are represented
as appropriate multiplication and differentiation operators:
\begin{equation}
\hat{\Omega}\psi(\omega) = \omega\psi(\omega), \quad
\hat{T}\psi(\omega) = i(1 + \kappa\omega^2)\partial_\omega \psi(\omega).
\label{eq:3}
\end{equation}

One can check with a direct calculation that the operators defined by
\eqref{eq:3} satisfy commutation relations \eqref{eq:1}.

Although eigenvectors of the time operator $\hat{T}$ are not physical states, it
is possible to define states which are physical and closely resemble eigenstates
of $\hat{T}$. These are states of maximal localization around instances of time.
A state $\ket{\varphi_\tau^{ML}}_T$ of maximal localization around an instance
of time $\tau$ is a state for which the expectation value of $\hat{T}$ is equal
$\tau$ and the uncertainty of $\hat{T}$ is equal to the smallest possible value
$\Delta t_0 = \sqrt{\kappa}$:
\begin{equation}
{}_T\bra{\varphi_\tau^{ML}}\hat{T}\ket{\varphi_\tau^{ML}}_T = \tau, \quad
(\Delta t)_{\ket{\varphi_\tau^{ML}}_T} = \Delta t_0.
\label{eq:4}
\end{equation}

The conditions in \eqref{eq:4} uniquely determine the state
$\ket{\varphi_\tau^{ML}}_T$, which in frequency representation is given by the
formula
\begin{equation}
\varphi_\tau^{ML}(\omega) = \sqrt{\frac{2\sqrt{\kappa}}{\pi}} (1 + \kappa\omega^2)^{-1/2} \exp\left(-i\tau\frac{\arctan(\sqrt{\kappa}\omega)}{\sqrt{\kappa}}\right).
\end{equation}

Using the states of maximal localization, it is possible to define a
representation of the Hilbert space $\mathcal{H}_T$, which we will call a
continuous time representation. This representation is given by projecting
states $\ket{\psi}_T$ in $\mathcal{H}_T$ with states of maximal localization
\begin{equation}
\psi(\tau) = {}_T\braket{\varphi_\tau^{ML}|\psi}_T.
\label{eq:5}
\end{equation}

The received wave functions $\psi(\tau)$ describe the probability amplitude
for the system being maximally localized around the instance of time $\tau$.
In the limit $\kappa \to 0$, the wave function $\psi(\tau) =
{}_T\braket{\tau|\psi}_T$ of the ordinary time representation is recovered.
From \eqref{eq:5}, we receive the transformation of a state's wave function
in the frequency representation into its continuous time representation:
\begin{equation}
\psi(\tau) = \sqrt{\frac{2\sqrt{\kappa}}{\pi}} \int_{-\infty}^{+\infty} \frac{\dd\omega}{(1 + \kappa\omega^2)^{3/2}} \exp\left(i\tau\frac{\arctan(\sqrt{\kappa}\omega)}{\sqrt{\kappa}}\right) \psi(\omega).
\label{eq:8}
\end{equation}

The inverse transformation is given by the formula
\begin{equation}
\psi(\omega) = \frac{1}{\sqrt{8\pi\sqrt{\kappa}}} \int_{-\infty}^{+\infty} \vd{\tau} (1 + \kappa\omega^2)^{1/2} \exp\left(-i\tau\frac{\arctan(\sqrt{\kappa}\omega)}{\sqrt{\kappa}}\right) \psi(\tau).
\end{equation}

The operators $\hat{\Omega}$ and $\hat{T}$ in the continuous time representation
take the form
\begin{equation}
\hat{\Omega}\psi(\tau) = \frac{\tan(-i\sqrt\kappa\partial_\tau)}{\sqrt\kappa} \psi(\tau), \quad
\hat{T}\psi(\tau) = \left(\tau + i\kappa\frac{\tan(-i\sqrt\kappa\partial_\tau)}{\sqrt\kappa}\right)\psi(\tau).
\label{eq:13}
\end{equation}

\subsection{Discrete Time Representation}
\label{subsec:3.2}
The family of states $\{\ket{\varphi_\tau^{ML}}_T\}$ for $\tau \in \mathbb{R}$
forms an overcomplete set of vectors. We can choose from this set a smaller
countable set forming a basis in the Hilbert space $\mathcal{H}_T$. Using this
basis, we can construct another representation of $\mathcal{H}_T$ which we will
call a discrete time representation. For $\lambda \in [0,1)$, let us consider
the following set of vectors:
\begin{equation}
\{\ket{\varphi_{2\sqrt{\kappa}(\lambda+n)}^{ML}}_T \mid n \in \mathbb{Z}\}.
\label{eq:6}
\end{equation}

The vectors composing this set are linearly independent and the set itself is
complete. Thus, we receive a one-parameter family of bases of the Hilbert space
$\mathcal{H}_T$. These bases can be viewed as lattices with spacing
$2\sqrt\kappa = 2\Delta t_0$ shifted by $2\sqrt{\kappa}\lambda$. As we will see,
the resulting discrete time representations will describe time evolution on
lattices. The vectors from the set \eqref{eq:6} are not orthogonal. The scalar
product of two states of maximal localization is equal:
\begin{equation}
{}_T\braket{\varphi_{\tau'}^{ML}|\varphi_\tau^{ML}}_T =
    \frac{1}{\pi} \left( \frac{\tau - \tau'}{2\sqrt\kappa} - \left(\frac{\tau - \tau'}{2\sqrt\kappa}\right)^3 \right)^{-1} \sin\left(\frac{\tau - \tau'}{2\sqrt\kappa}\pi\right).
\end{equation}

From this, we can see that
\begin{equation}
{}_T\braket{\varphi_{2\sqrt{\kappa}(\lambda+m)}^{ML}|\varphi_{2\sqrt{\kappa}(\lambda+n)}^{ML}}_T = \begin{cases}
1 & \text{for $m = n$} \\
\tfrac{1}{2} & \text{for $m = n \pm 1$} \\
0 & \text{otherwise}
\end{cases}
\end{equation}
so that the vectors from the set \eqref{eq:6} are orthonormal except for
neighboring vectors.

Using a basis \eqref{eq:6}, we can construct a discrete time representation of
the Hilbert space $\mathcal{H}_T$. Instead of expanding a state $\ket{\psi}_T$
in this basis and taking the sequence of coefficients of this expansion as the
representation of the vector $\ket{\psi}_T$, we will represent the state
$\ket{\psi}_T$ with the following sequence:
\begin{equation}
\psi_n = {}_T\braket{\varphi_{2\sqrt{\kappa}(\lambda+n)}^{ML}|\psi}_T.
\end{equation}

From this formula, we receive the transformation of a state's wave function in
the frequency representation into its discrete time representation:
\begin{equation}
\psi_n = \sqrt{\frac{2\sqrt{\kappa}}{\pi}} \int_{-\infty}^{+\infty}
    \frac{\dd\omega}{(1 + \kappa\omega^2)^{3/2}} e^{2i(\lambda+n)\arctan(\sqrt{\kappa}\omega)} \psi(\omega).
\label{eq:7}
\end{equation}

To calculate the inverse transform, let us rewrite the above formula by changing
the variable under the integral sign:
\begin{equation}
\psi_n = \frac{1}{2\pi} \int_{-\pi}^{+\pi} \sqrt{\frac{2\pi}{\sqrt\kappa}} \psi\left(\frac{1}{\sqrt\kappa}\tan\frac{x}{2}\right) \frac{1}{(1 + \tan^2\frac{x}{2})^{1/2}} e^{i\lambda x} e^{inx} \ud{x}.
\end{equation}

From this, we can see that $\psi_n$ are Fourier coefficients in the expansion of
the function
\begin{equation}
f(x) = \sqrt{\frac{2\pi}{\sqrt\kappa}} \psi\left(\frac{1}{\sqrt\kappa}\tan\frac{x}{2}\right) \frac{1}{(1 + \tan^2\frac{x}{2})^{1/2}} e^{i\lambda x}
\end{equation}
into the Fourier series:
\begin{equation}
f(x) = \sum_{n=-\infty}^\infty \psi_n e^{-inx}.
\end{equation}

The function $f$ is square integrable, which follows from the square
integrability of $\psi$ with respect to the measure $\dd{\mu}(\omega) =
\frac{\dd\omega}{1 + \kappa\omega^2}$. Therefore, by virtue of Carleson's
theorem, the above Fourier series converges for almost all $x$, where the
infinite sum $\textstyle \sum_{n=-\infty}^\infty$ is viewed as the limit
$\textstyle \lim_{N \to \infty} \sum_{n=-N}^N$. We can then change the variable
again to receive the following inverse transform of \eqref{eq:7}:
\begin{equation}
\psi(\omega) = \sqrt{\frac{\sqrt\kappa}{2\pi}} \sum_{n=-\infty}^\infty
    (1 + \kappa\omega^2)^{1/2} e^{-2i(\lambda+n)\arctan(\sqrt{\kappa}\omega)} \psi_n.
\label{eq:9}
\end{equation}

The transformation of a state's wave function in the continuous time
representation into its discrete time representation is the following:
\begin{equation}
\psi_n = \psi\left(2\sqrt{\kappa}(\lambda+n)\right).
\label{eq:10}
\end{equation}

To calculate the inverse transform, we combine transformations \eqref{eq:8} and
\eqref{eq:9} receiving
\begin{linenomath}
\begin{align}
\psi(\tau) & = \frac{\sqrt\kappa}{\pi} \int_{-\infty}^{+\infty} \sum_{n=-\infty}^\infty
    \psi_n e^{i(\tau/\sqrt{\kappa} - 2(\lambda+n))\arctan(\sqrt{\kappa}\omega)} \frac{\dd\omega}{1 + \kappa\omega^2} \nonumber \\
& = \frac{1}{\pi} \sum_{n=-\infty}^\infty \psi_n \int_{-\pi/2}^{+\pi/2} e^{i(\tau/\sqrt{\kappa} - 2(\lambda+n))x} \ud{x}.
\end{align}
\end{linenomath}

Calculating the integral in the above formula, we receive the inverse transform
of \eqref{eq:10}:
\begin{equation}
\psi(\tau) = \sum_{n=-\infty}^\infty \psi_n \sinc\left(\frac{\tau - 2\sqrt{\kappa}(\lambda+n)}{2\sqrt\kappa}\right),
\label{eq:11}
\end{equation}
where $\sinc x = \frac{\sin(\pi x)}{\pi x}$. The transformations \eqref{eq:10}
and \eqref{eq:11} might seem surprising because they allow us to reproduce the
function $\psi(\tau)$ from the knowledge of its values at a countable set of
points. This is, however, nothing strange because from \eqref{eq:8} after
changing the variable under the integral, we can see that $\psi(\tau)$ is a
Fourier transform of a square integrable function with compact support. Thus,
by virtue of Paley--Wiener theorem, $\psi(\tau)$ extends to an entire function
of exponential type and entire functions are completely determined by their
values at a countable set of points.

In what follows, we will derive the representation of the frequency operator
$\hat{\Omega}$ in the discrete time representation. Let us define a discrete
derivative with the formula
\begin{equation}
D_n\psi_n = \frac{\psi_{n+1} - \psi_{n-1}}{4\Delta t_0}.
\end{equation}

Using \eqref{eq:7} we calculate
\vspace{-6pt}
\begin{adjustwidth}{-\extralength}{0cm}
\begin{linenomath}
\begin{align}
-iD_n\psi_n & = -i\frac{\psi_{n+1} - \psi_{n-1}}{4\sqrt\kappa} \nonumber \\
& = \frac{-i}{4\sqrt\kappa} \sqrt{\frac{2\sqrt{\kappa}}{\pi}} \int_{-\infty}^{+\infty}
    \frac{\dd\omega}{(1 + \kappa\omega^2)^{3/2}}
    \Bigl(e^{2i(\lambda+n+1)\arctan(\sqrt{\kappa}\omega)} - e^{2i(\lambda+n-1)\arctan(\sqrt{\kappa}\omega)}\Bigr) \psi(\omega) \nonumber \\
& = \frac{-i}{4\sqrt\kappa} \sqrt{\frac{2\sqrt{\kappa}}{\pi}} \int_{-\infty}^{+\infty}
    \frac{\dd\omega}{(1 + \kappa\omega^2)^{3/2}} e^{2i(\lambda+n)\arctan(\sqrt{\kappa}\omega)}
    \Bigl(e^{2i\arctan(\sqrt{\kappa}\omega)} - e^{-2i\arctan(\sqrt{\kappa}\omega)}\Bigr) \psi(\omega) \displaybreak[0] \nonumber \\
& = \sqrt{\frac{2\sqrt{\kappa}}{\pi}} \int_{-\infty}^{+\infty}
    \frac{\dd\omega}{(1 + \kappa\omega^2)^{3/2}} e^{2i(\lambda+n)\arctan(\sqrt{\kappa}\omega)} \frac{1}{2\sqrt\kappa} \sin\left(2\arctan(\sqrt{\kappa}\omega)\right) \psi(\omega) \nonumber \\
& = \sqrt{\frac{2\sqrt{\kappa}}{\pi}} \int_{-\infty}^{+\infty}
    \frac{\dd\omega}{(1 + \kappa\omega^2)^{3/2}} e^{2i(\lambda+n)\arctan(\sqrt{\kappa}\omega)} \frac{\omega}{1 + \kappa\omega^2} \psi(\omega).
\end{align}
\end{linenomath}
\end{adjustwidth}

From this we get
\begin{equation}
(-iD_n)^k \psi_n = \sqrt{\frac{2\sqrt{\kappa}}{\pi}} \int_{-\infty}^{+\infty}
    \frac{\dd\omega}{(1 + \kappa\omega^2)^{3/2}} e^{2i(\lambda+n)\arctan(\sqrt{\kappa}\omega)} \left(\frac{\omega}{1 + \kappa\omega^2}\right)^k \psi(\omega)
\end{equation}
and consequently
\begin{equation}
\frac{1}{\sqrt\kappa} f(-i\sqrt{\kappa}D_n) \psi_n = \sqrt{\frac{2\sqrt{\kappa}}{\pi}} \int_{-\infty}^{+\infty}
    \frac{\dd\omega}{(1 + \kappa\omega^2)^{3/2}} e^{2i(\lambda+n)\arctan(\sqrt{\kappa}\omega)} \omega \psi(\omega),
\label{eq:12}
\end{equation}
where $f(x) = \frac{2x}{1 + \sqrt{1 - 4x^2}}$ is a function with an inverse
$f^{-1}(x) = \frac{x}{1 + x^2}$ and we used the expansion of this function in
a Taylor series
\begin{equation}
f(x) = x + x^3 + 2x^5 + 5x^7 + \dotsb.
\end{equation}

From \eqref{eq:12} and the fact that in the frequency representation the
operator $\hat{\Omega}$ is a multiplication operator (see \eqref{eq:3}) we
receive the discrete time representation of the frequency operator
\begin{equation}
\hat{\Omega}\psi_n = \frac{f(-i\sqrt{\kappa}D_n)}{\sqrt\kappa} \psi_n.
\label{eq:14}
\end{equation}

Using this result and Formulas \eqref{eq:3} and \eqref{eq:7} we can also receive
the discrete time representation of the time operator
\begin{equation}
\hat{T}\psi_n = \left(2\sqrt{\kappa}(\lambda+n) + i\kappa\frac{f(-i\sqrt{\kappa}D_n)}{\sqrt\kappa}\right)\psi_n.
\end{equation}

\subsection{Time Evolution with a Minimal Time Scale}
\label{subsec:3.3}
The introduction of a minimal time scale into the formalism will change the
time evolution of the system. Since the time operator $\hat{T}$ is no longer
self-adjoint, its eigenvectors are not physical states and it is meaningless to
talk about the system at particular instances of time. We can only consider the
system localized around instances of time with a finite precision. For this
reason, as indicators of time of the evolving system, we will take states of
maximal localization around instances of time $\ket{\varphi_\tau^{ML}}_T$
instead of eigenstates $\ket{t}_T$ of the time operator $\hat{T}$.

The state $\ket{\psi(\tau)}_S$ of the system $S$ maximally localized around an
instance of time $\tau$ can be obtained via projection of the physical state
$\ketket{\Psi}$ with the maximal localization state $\ket{\varphi_\tau^{ML}}_T$:
\begin{equation}
\ket{\psi(\tau)}_S = {}_T\braketket{\varphi_\tau^{ML}|\Psi}.
\end{equation}

To receive a modified Schr\"odinger equation governing the time evolution of
states $\ket{\psi(\tau)}_S$, we write the constraint equation
$\hat{J}\ketket{\Psi} = 0$ in the continuous or discrete time representation in
$\mathcal{H}_T$. Before doing this, first let us use Theorem~\ref{thm:1} from
Appendix \ref{sec:AA} with the function
$f(x) = \frac{1}{\sqrt\kappa}\arctan(\sqrt{\kappa}x)$ to rewrite the constraint
equation
\begin{equation}
\left(-\hat{\Omega} \otimes \hat{1}_S\right) \ketket{\Psi} = \left(\hat{1}_T \otimes \hat{H}_S/\hbar\right) \ketket{\Psi}
\end{equation}
in a different form
\begin{linenomath}
\begin{gather}
\frac{1}{\sqrt\kappa}\arctan\left(-\sqrt{\kappa}\,\hat{\Omega} \otimes \hat{1}_S\right) \ketket{\Psi} =
    \frac{1}{\sqrt\kappa}\arctan\left(\sqrt{\kappa}\,\hat{1}_T \otimes \hat{H}_S/\hbar\right) \ketket{\Psi} \nonumber \\
\Big\Updownarrow \\
\left(\frac{1}{\sqrt\kappa}\arctan\left(-\sqrt{\kappa}\,\hat{\Omega}\right) \otimes \hat{1}_S\right) \ketket{\Psi} =
    \left(\hat{1}_T \otimes \frac{1}{\sqrt\kappa}\arctan\left(\sqrt{\kappa}\,\hat{H}_S/\hbar\right)\right) \ketket{\Psi}. \nonumber
\end{gather}
\end{linenomath}

In the continuous time representation, the constraint equation takes the form
\begin{linenomath}
\begin{gather}
{}_T\bra{\varphi_\tau^{ML}}\hat{J}\ketket{\Psi} = 0 \nonumber \\
\Big\Updownarrow \\
{}_T\bra{\varphi_\tau^{ML}}\left(\frac{1}{\sqrt\kappa}\arctan\left(-\sqrt{\kappa}\,\hat{\Omega}\right) \otimes \hat{1}_S\right) \ketket{\Psi} =
    {}_T\bra{\varphi_\tau^{ML}}\left(\hat{1}_T \otimes \frac{1}{\sqrt\kappa}\arctan\left(\sqrt{\kappa}\,\hat{H}_S/\hbar\right)\right) \ketket{\Psi} \nonumber
\end{gather}
\end{linenomath}
from which and \eqref{eq:13} we receive the modified Schr\"odinger equation
\begin{equation}
i\frac{\partial}{\partial\tau} \ket{\psi(\tau)}_S = \frac{1}{\sqrt\kappa}\arctan\left(\sqrt\kappa\,\hat{H}_S/\hbar\right) \ket{\psi(\tau)}_S.
\label{eq:15}
\end{equation}

Similarly, by writing the constraint equation in the discrete time
representation and using~\eqref{eq:14}, we receive the discrete Schr\"odinger
equation
\begin{equation}
i\hbar D_n\ket{\psi_n}_S = \hat{H}_S \left(\hat{1} + \kappa(\hat{H}_S/\hbar)^2\right)^{-1} \ket{\psi_n}_S.
\label{eq:16}
\end{equation}

The solution of \eqref{eq:15} subject to an initial condition
$\ket{\psi(\tau = 0)}_S = \ket{\psi_0}_S$ is of the form
\begin{equation}
\ket{\psi(\tau)}_S = \exp\left(-i\tau\frac{1}{\sqrt\kappa}\arctan(\sqrt\kappa \hat{H}_S/\hbar)\right) \ket{\psi_0}_S
\label{eq:17}
\end{equation}
and the solution of \eqref{eq:16} subject to an initial condition
$\ket{\psi_{n=0}}_S = \ket{\psi_0}_S$ is of the form
\begin{equation}
\ket{\psi_n}_S = \exp\left(-2in\arctan(\sqrt\kappa \hat{H}_S/\hbar)\right) \ket{\psi_0}_S.
\end{equation}

Equations \eqref{eq:15} and \eqref{eq:16} are equivalent and describe the same
time evolution of the system. Knowing the solution of either of these equations,
we can reconstruct the solution of the other equation by applying the transform
\eqref{eq:10} or \eqref{eq:11}.

\section{Examples}
\label{sec:4}

\subsection{Spin-$1/2$ in an External Magnetic Field}
\label{subsec:4.1}
In this section, we will investigate how the minimal time scale influences the
time evolution of a spin-$1/2$ in an external constant magnetic field. The
Hilbert space of the system is $\mathcal{H}_S = \mathbb{C}^2$ and the
Hamiltonian of the system is
\begin{equation}
\hat{H}_S = -\gamma B_0 \hat{S}_z = -\gamma B_0 \frac{\hbar}{2} \sigma_z,
\end{equation}
where $\gamma$ is the gyromagnetic ratio; $B_0$ is the strength of the magnetic
field, which we assume is directed along the $z$-axis; and $\hat{S}_z$ is the
$z$-component of the spin operator expressed by the Pauli matrix
\begin{equation}
\sigma_z = \begin{bmatrix}
1 & 0 \\
0 & -1
\end{bmatrix}.
\end{equation}

Let us assume that the system is initially in an arbitrary state:
\begin{equation}
\ket{\psi}_S = \ket{\theta,\varphi}_S = \cos(\theta/2) e^{-i\varphi/2} \ket{\uparrow}_S + \sin(\theta/2) e^{i\varphi/2} \ket{\downarrow}_S,
\end{equation}
where $\theta,\varphi \in \mathbb{R}$ and $\ket{\uparrow}_S$,
$\ket{\downarrow}_S$ are spin-up and spin-down eigenstates of the spin operator
$\hat{S}_z$. In the case without a minimal time scale ($\kappa = 0$), the state
$\ket{\psi}_S$ will evolve in time according to the formula
\begin{linenomath}
\begin{align}
\ket{\psi(t)}_S & = e^{-it\hat{H}_S/\hbar} \ket{\psi}_S = e^{-it\omega_0\sigma_z/2} \ket{\psi}_S \nonumber \\
& = \cos(\theta/2) e^{-i(\varphi + \omega_0 t)/2} \ket{\uparrow}_S + \sin(\theta/2) e^{i(\varphi + \omega_0 t)/2} \ket{\downarrow}_S \nonumber \\
& = \ket{\theta,\varphi + \omega_0 t}_S,
\end{align}
\end{linenomath}
where $\omega_0 = -\gamma B_0$. The spin is precessing around the $z$-axis with
the frequency $\omega_0$. This is the usual Larmour precession.

Let us now consider the case of a non-zero minimal time scale ($\kappa > 0$).
The state $\ket{\psi}_S$ will evolve in time according to the formula
\eqref{eq:17}
\begin{equation}
\ket{\psi(\tau)}_S = \exp\left(-i\tau\frac{1}{\sqrt\kappa}\arctan(\sqrt\kappa \omega_0 \sigma_z/2)\right) \ket{\psi}_S.
\end{equation}

Using the fact that $\sigma_z^2 = I$, we can calculate explicitly $\arctan$ of
the matrix $\sigma_z$ receiving
\begin{linenomath}
\begin{align}
\ket{\psi(\tau)}_S & = \exp\left(-i\tau\frac{1}{\sqrt\kappa}\arctan(\sqrt\kappa \omega_0/2)\sigma_z\right) \ket{\psi}_S \nonumber \\
& = \exp\left(-i\tau \omega_\kappa \sigma_z/2\right) \ket{\psi}_S \nonumber \\
& = \cos(\theta/2) e^{-i(\varphi + \omega_\kappa \tau)/2} \ket{\uparrow}_S + \sin(\theta/2) e^{i(\varphi + \omega_\kappa \tau)/2} \ket{\downarrow}_S \nonumber \\
& = \ket{\theta,\varphi + \omega_\kappa \tau}_S,
\label{eq:24}
\end{align}
\end{linenomath}
where $\omega_\kappa = \frac{2}{\sqrt\kappa}\arctan(\sqrt\kappa \omega_0/2)$.
The spin is now precessing around the $z$-axis with the smaller frequency
$\omega_\kappa$. Notice, moreover, that since $\arctan$ is a bounded function,
there is an upper limit for the frequency with which a spin can precess. The
frequency of precession is necessarily smaller than $\pi/\sqrt\kappa$ and this
upper limit is independent on the strength of the magnetic field. This
phenomenon results from the existence of the fundamental limit for the precision
with which we can measure time.

\subsection{Free Particle}
\label{subsec:4.2}
In this section, we will investigate how the minimal time scale influences the
time evolution of a free particle moving on a line. The Hilbert space of the
system is $\mathcal{H}_S~=~L^2(\mathbb{R},\dd{x})$ and the Hamiltonian of the
system is
\begin{equation}
\hat{H}_S = \frac{1}{2m} \hat{p}^2,
\end{equation}
where $m$ is the mass of a particle. Let us assume that the system is initially
in an arbitrary state written in the position representation as
\begin{equation}
\psi(x) = \frac{1}{\sqrt{2\pi\hbar}} \int_{-\infty}^{+\infty} f(p) e^{\frac{i}{\hbar}px} \ud{p},
\end{equation}
where $f$ is an arbitrary function such that $\psi$ will be in the domains of
the position and momentum operators $\hat{x}$, $\hat{p}$ and their squares
$\hat{x}^2$, $\hat{p}^2$. It is not difficult to see that the function
\begin{equation}
\psi(\tau,x) = \frac{1}{\sqrt{2\pi\hbar}} \int_{-\infty}^{+\infty} f(p) e^{\frac{i}{\hbar}(px - E(p)\tau)} \ud{p},
\label{eq:18}
\end{equation}
where
\begin{equation}
E(p) = \frac{\hbar}{\sqrt\kappa}\arctan\left(\frac{\sqrt{\kappa}p^2}{2m\hbar}\right)
\end{equation}
is a solution to the modified Schr\"odinger Equation \eqref{eq:15} satisfying
the initial condition $\psi(\tau = 0,x) = \psi(x)$.

In accordance with \eqref{eq:15}, the effective Hamiltonian describing the time
evolution of the system is
\begin{equation}
\hat{H}_{\mathrm{eff}} = \frac{\hbar}{\sqrt\kappa}\arctan\left(\sqrt\kappa\,\hat{H}_S/\hbar\right)
= \frac{\hbar}{\sqrt\kappa}\arctan\left(\frac{\sqrt\kappa\hat{p}^2}{2m\hbar}\right).
\end{equation}

By the first equation of Hamilton's equations in classical Hamiltonian
mechanics
\begin{equation}
\dot{q} = \frac{\partial H}{\partial p}, \quad \dot{p} = -\frac{\partial H}{\partial q}
\end{equation}
we can define a velocity operator by the formula
\begin{equation}
\hat{v} = \frac{\hat{p}}{m} \left(\hat{1} + \frac{\kappa\hat{p}^4}{4m^2\hbar^2}\right)^{-1}.
\end{equation}

The expectation value of the momentum operator $\hat{p}$ in the state
$\psi(\tau,x)$ is equal
\begin{equation}
\braket{\hat{p}}_{\psi(\tau)} = \int_{-\infty}^{+\infty} \overline{\tilde{\psi}(\tau,p)} p \tilde{\psi}(\tau,p) \ud{p}
= \int_{-\infty}^{+\infty} p \abs{f(p)}^2 \ud{p} = \braket{\hat{p}}_{\psi(0)},
\end{equation}
where $\tilde{\psi}(\tau,p) = f(p) e^{-\frac{i}{\hbar}E(p)\tau}$ is the Fourier
transform of $\psi(\tau,x)$. Notice that the expectation value of the momentum
is the same as in the $\kappa = 0$ case and is independent of~time.

The expectation value of the position operator $\hat{x}$ in the state
$\psi(\tau,x)$ is equal to
\begin{linenomath}
\begin{align}
\braket{\hat{x}}_{\psi(\tau)} & = \int_{-\infty}^{+\infty} \overline{\tilde{\psi}(\tau,p)} i\hbar\partial_p \tilde{\psi}(\tau,p) \ud{p} \nonumber \\
& = i\hbar \int_{-\infty}^{+\infty} \overline{f(p)} f'(p) \ud{p}
    + \tau \int_{-\infty}^{+\infty} \abs{f(p)}^2 E'(p) \ud{p} \nonumber \\
& = \braket{\hat{x}}_{\psi(0)}
    + \frac{\tau}{m} \int_{-\infty}^{+\infty} \abs{f(p)}^2 \frac{p}{1 + \kappa p^4/4m^2\hbar^2} \ud{p} \nonumber \\
& = \braket{\hat{x}}_{\psi(0)} + \tau \braket{\hat{v}}_{\psi(0)},
\end{align}
\end{linenomath}
which shows that indeed $\hat{v}$ can be interpreted as the velocity operator.

The expectation value of the velocity operator $\hat{v}$ in the state
$\psi(\tau,x)$ is equal to
\begin{equation}
\braket{\hat{v}}_{\psi(\tau)} = \frac{1}{m} \int_{-\infty}^{+\infty} \abs{f(p)}^2 \frac{p}{1 + \kappa p^4/4m^2\hbar^2} \ud{p}
\end{equation}
and is independent of time. Notice that the expectation value of the velocity
is smaller than in the $\kappa = 0$ case, where it is equal to the expectation
value of the momentum divided by the mass $m$. Since the state $\psi(\tau,x)$ is
normalized to unity and the function $\frac{p/m}{1 + \kappa p^4/4m^2\hbar^2}$
is bounded with a maximum value equal
\begin{equation}
v_{\mathrm{max}} = \sqrt{\frac{3\sqrt{3}\hbar}{8m\sqrt{\kappa}}}
\end{equation}
this value will be the upper limit on the speed of propagation of wave packets.
Notice that $v_{\mathrm{max}}$ does not depend on the initial state $\psi(x)$,
so in particular on the expectation value of the momentum,
but it depends on the mass $m$ of the particle. The heavier the particle, the
smaller this speed limit is. The existence of the upper limit with which wave
packets can propagate is another consequence of the fundamental limit for the
precision with which we can measure time.

As an example, let us calculate $v_{\mathrm{max}}$ for the proton assuming that
$\Delta t_0 = \sqrt{\kappa}$ is equal to the Planck's time $t_p$. We obtain the
value
\begin{equation}
v_{\mathrm{max}} \approx 8.72 \cdot 10^{17}\,\frac{\mathrm{m}}{\mathrm{s}},
\end{equation}
which is much bigger than the speed of light. It should be noted that we were
not considering the relativistic mechanics, which is the reason we received
a speed limit bigger than the speed of light. However, for an object with high
enough mass, $v_{\mathrm{max}}$ can be much smaller than $c$. For example, when
$m$ is equal to the Planck's mass $m_p$, then
\begin{equation}
v_{\mathrm{max}} = \sqrt{\frac{3\sqrt{3}}{8}}c \approx 0.8c
\end{equation}
and when $m = 150\sqrt{3}m_p \approx 5.65\,\text{mg}$, then $v_{\mathrm{max}} =
0.05c$.

The minimal time scale will also influence the wave packet spreading.
Calculating the uncertainty in position for the state \eqref{eq:18}, we obtain
\begin{equation}
\Delta x_{\psi(\tau)} = \sqrt{(\Delta x_{\psi(0)})^2
    + \tau\bigl(\braket{\hat{v}\hat{x} + \hat{x}\hat{v}}_{\psi(0)} - 2\braket{\hat{x}}_{\psi(0)}\braket{\hat{v}}_{\psi(0)}\bigr)
    + \tau^2 (\Delta v_{\psi(0)})^2},
\label{eq:19}
\end{equation}
which is a similar looking formula as in the $\kappa = 0$ case except that the
velocity operator $\hat{v}$ is expressed by a different formula. Therefore,
the uncertainty in position will increase in time, which is interpreted as the
spreading of the wave packet evolving in time. However, $\Delta x_{\psi(\tau)}$
will change in time in a slightly different way than in the $\kappa = 0$ case.
Formula \eqref{eq:19} simplifies for states \eqref{eq:18} for which the function
$f$ is real-valued. Then, by integration by parts, it is easy to see that
$\braket{\hat{x}}_{\psi(0)} = 0$ and $\braket{\hat{v}\hat{x} +
\hat{x}\hat{v}}_{\psi(0)} = 0$ from which we get
\begin{equation}
\Delta x_{\psi(\tau)} = \sqrt{(\Delta x_{\psi(0)})^2 + \tau^2 (\Delta v_{\psi(0)})^2}.
\end{equation}

As an example, we can take the state \eqref{eq:18} in the form of a Gaussian
wave packet by taking the function $f$ in the form of a Gaussian function:
\begin{equation}
f(p) = \frac{1}{(2\pi)^{1/4}(\Delta p)^{1/2}} \exp\left(-\frac{(p - p_0)^2}{4(\Delta p)^2}\right).
\end{equation}

\begin{figure}
\begin{tabular}{cc}
\includegraphics{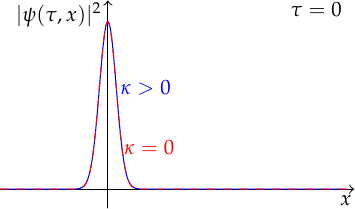}
%\begin{asy}
%usepackage("mathpazo");
%import graph;
%size(170,100,IgnoreAspect);
%defaultpen(fontsize(10pt));

%real kappa = 0.005;
%real p0 = 3;
%real Dx = 1/sqrt(2);
%real Dp = 1/(2*Dx);
%real t = 0;

%real AbsPhi2(real x) { return 1/sqrt(2*pi*(Dx^2+t^2*Dp^2)) * exp(-(x-p0*t)^2/(2*Dx^2+2*t^2*Dp^2)); }

%real E(real p) { return 1/sqrt(kappa) * atan(sqrt(kappa)*p^2/2); }
%real f(real p) { return 1/sqrt(sqrt(2*pi)*Dp) * exp(-(p-p0)^2/(4*Dp^2)); }
%real RePsi(real x) { return 1/sqrt(2*pi) * simpson(new real(real p) { return f(p)*cos(p*x-E(p)*t); }, -100, 100); }
%real ImPsi(real x) { return 1/sqrt(2*pi) * simpson(new real(real p) { return f(p)*cos(p*x-E(p)*t-pi/2); }, -100, 100); }
%real AbsPsi2(real x) { return RePsi(x)^2 + ImPsi(x)^2; }

%draw(graph(AbsPsi2,-9,20,operator ..),blue);
%draw(graph(AbsPhi2,-9,20,operator ..),red+dashed);
%xaxis("$x$",Arrow(TeXHead));
%yaxis("$|\psi(\tau,x)|^2$",Arrow(TeXHead),autorotate=false);

%label("$\tau = 0$",(20,1/(sqrt(2*pi)*Dx)),NW);
%label("$\kappa = 0$",(1,0.1),NE,red);
%label("$\kappa > 0$",(0.7,0.3),NE,blue);
%\end{asy}
&
\includegraphics{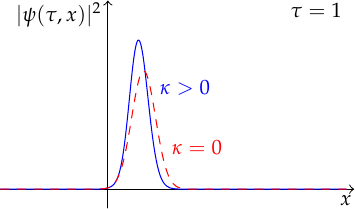}
%\begin{asy}
%usepackage("mathpazo");
%import graph;
%size(170,100,IgnoreAspect);
%defaultpen(fontsize(10pt));

%real kappa = 0.005;
%real p0 = 3;
%real Dx = 1/sqrt(2);
%real Dp = 1/(2*Dx);
%real t = 1;

%real AbsPhi2(real x) { return 1/sqrt(2*pi*(Dx^2+t^2*Dp^2)) * exp(-(x-p0*t)^2/(2*Dx^2+2*t^2*Dp^2)); }

%real E(real p) { return 1/sqrt(kappa) * atan(sqrt(kappa)*p^2/2); }
%real f(real p) { return 1/sqrt(sqrt(2*pi)*Dp) * exp(-(p-p0)^2/(4*Dp^2)); }
%real RePsi(real x) { return 1/sqrt(2*pi) * simpson(new real(real p) { return f(p)*cos(p*x-E(p)*t); }, -100, 100); }
%real ImPsi(real x) { return 1/sqrt(2*pi) * simpson(new real(real p) { return f(p)*cos(p*x-E(p)*t-pi/2); }, -100, 100); }
%real AbsPsi2(real x) { return RePsi(x)^2 + ImPsi(x)^2; }

%draw(graph(AbsPsi2,-9,20,operator ..),blue);
%draw(graph(AbsPhi2,-9,20,operator ..),red+dashed);
%xaxis("$x$",Arrow(TeXHead));
%yaxis("$|\psi(\tau,x)|^2$",Arrow(TeXHead),autorotate=false);

%label("$\tau = 1$",(20,1/(sqrt(2*pi)*Dx)),NW);
%label("$\kappa = 0$",(5,0.1),NE,red);
%label("$\kappa > 0$",(4,0.3),NE,blue);
%\end{asy}
\\
(\textbf{a}) & (\textbf{b}) \\
\includegraphics{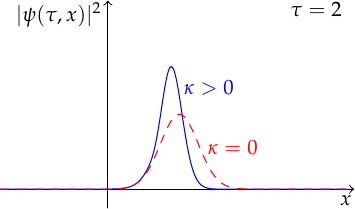}
%\begin{asy}
%usepackage("mathpazo");
%import graph;
%size(170,100,IgnoreAspect);
%defaultpen(fontsize(10pt));

%real kappa = 0.005;
%real p0 = 3;
%real Dx = 1/sqrt(2);
%real Dp = 1/(2*Dx);
%real t = 2;

%real AbsPhi2(real x) { return 1/sqrt(2*pi*(Dx^2+t^2*Dp^2)) * exp(-(x-p0*t)^2/(2*Dx^2+2*t^2*Dp^2)); }

%real E(real p) { return 1/sqrt(kappa) * atan(sqrt(kappa)*p^2/2); }
%real f(real p) { return 1/sqrt(sqrt(2*pi)*Dp) * exp(-(p-p0)^2/(4*Dp^2)); }
%real RePsi(real x) { return 1/sqrt(2*pi) * simpson(new real(real p) { return f(p)*cos(p*x-E(p)*t); }, -100, 100); }
%real ImPsi(real x) { return 1/sqrt(2*pi) * simpson(new real(real p) { return f(p)*cos(p*x-E(p)*t-pi/2); }, -100, 100); }
%real AbsPsi2(real x) { return RePsi(x)^2 + ImPsi(x)^2; }

%draw(graph(AbsPsi2,-9,20,operator ..),blue);
%draw(graph(AbsPhi2,-9,20,operator ..),red+dashed);
%xaxis("$x$",Arrow(TeXHead));
%yaxis("$|\psi(\tau,x)|^2$",Arrow(TeXHead),autorotate=false);

%label("$\tau = 2$",(20,1/(sqrt(2*pi)*Dx)),NW);
%label("$\kappa = 0$",(8,0.1),NE,red);
%label("$\kappa > 0$",(6,0.3),NE,blue);
%\end{asy}
&
\includegraphics{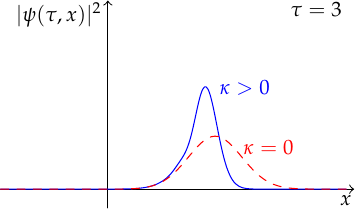}
%\begin{asy}
%usepackage("mathpazo");
%import graph;
%size(170,100,IgnoreAspect);
%defaultpen(fontsize(10pt));

%real kappa = 0.005;
%real p0 = 3;
%real Dx = 1/sqrt(2);
%real Dp = 1/(2*Dx);
%real t = 3;

%real AbsPhi2(real x) { return 1/sqrt(2*pi*(Dx^2+t^2*Dp^2)) * exp(-(x-p0*t)^2/(2*Dx^2+2*t^2*Dp^2)); }

%real E(real p) { return 1/sqrt(kappa) * atan(sqrt(kappa)*p^2/2); }
%real f(real p) { return 1/sqrt(sqrt(2*pi)*Dp) * exp(-(p-p0)^2/(4*Dp^2)); }
%real RePsi(real x) { return 1/sqrt(2*pi) * simpson(new real(real p) { return f(p)*cos(p*x-E(p)*t); }, -100, 100); }
%real ImPsi(real x) { return 1/sqrt(2*pi) * simpson(new real(real p) { return f(p)*cos(p*x-E(p)*t-pi/2); }, -100, 100); }
%real AbsPsi2(real x) { return RePsi(x)^2 + ImPsi(x)^2; }

%draw(graph(AbsPsi2,-9,20,operator ..),blue);
%draw(graph(AbsPhi2,-9,20,operator ..),red+dashed);
%xaxis("$x$",Arrow(TeXHead));
%yaxis("$|\psi(\tau,x)|^2$",Arrow(TeXHead),autorotate=false);

%label("$\tau = 3$",(20,1/(sqrt(2*pi)*Dx)),NW);
%label("$\kappa = 0$",(11,0.1),NE,red);
%label("$\kappa > 0$",(9,0.3),NE,blue);
%\end{asy}
\\
(\textbf{c}) & (\textbf{d})
\end{tabular}
\caption{(\textbf{a}--\textbf{d}) Plots of the probability density of the
Gaussian wave packet $\psi(\tau,x)$ with $\Delta p = 1/\sqrt{2}$ and $p_0 = 3$
for particular instances of time $\tau$. Mass $m = 1$, $\hbar = 1$,
$\kappa = 0.005$ (blue plots).}
\label{fig:2}
\end{figure}

\begin{figure}[H]
\begin{tabular}{cc}
\includegraphics{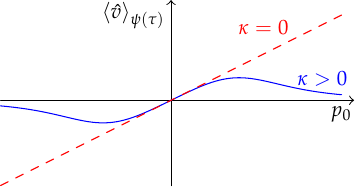}
%\begin{asy}
%usepackage("mathpazo");
%import graph;
%size(170,0);
%defaultpen(fontsize(10pt));

%real m = 2;
%real Dp = 1/sqrt(2);
%real kappa = 0.1;

%real v(real p0) { return 1/(sqrt(2*pi)*Dp) * simpson(new real(real p) { return 1/m*p/(1+kappa*p^4/(4*m^2)) * exp(-(p-p0)^2/(2*Dp^2)); }, -10, 10); }

%draw(graph(v,-7,7,operator ..),blue);
%draw((-7,-7/m)--(7,7/m),red+dashed);
%xaxis("$p_0$",Arrow(TeXHead));
%yaxis("$\langle\hat{v}\rangle_{\psi(\tau)}$",Arrow(TeXHead),autorotate=false);
%label("$\kappa = 0$",(5,5/m),NW,red);
%label("$\kappa > 0$",(5,0.4),NE,blue);
%\end{asy}
&
\includegraphics{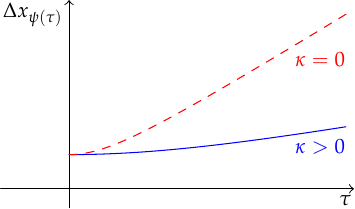}
%\begin{asy}
%usepackage("mathpazo");
%import graph;
%size(170,100,IgnoreAspect);
%defaultpen(fontsize(10pt));

%real m = 2;
%real p0 = 3;
%real Dx = 1/sqrt(2);
%real Dp = 1/(2*Dx);
%real kappa = 0.1;

%real v = 1/(sqrt(2*pi)*Dp) * simpson(new real(real p) { return 1/m*p/(1+kappa*p^4/(4*m^2)) * exp(-(p-p0)^2/(2*Dp^2)); }, -10, 10);
%real v2 = 1/(sqrt(2*pi)*Dp) * simpson(new real(real p) { return (1/m*p/(1+kappa*p^4/(4*m^2)))^2 * exp(-(p-p0)^2/(2*Dp^2)); }, -10, 10);
%real Dv2 = v2 - v^2;
%real Dx(real t) { return sqrt(Dx^2 + t^2 * Dv2); }
%real Dx0(real t) { return sqrt(Dx^2 + t^2 * (Dp/m)^2); }

%draw(graph(Dx,0,10,operator ..),blue);
%draw(graph(Dx0,0,10,operator ..),red+dashed);
%xaxis("$\tau$",Arrow(TeXHead));
%yaxis("$\Delta x_{\psi(\tau)}$",Arrow(TeXHead),autorotate=false);
%label("$\kappa = 0$",(8,Dx0(8)),SE,red);
%label("$\kappa > 0$",(8,Dx(8)),SE,blue);
%\end{asy}
\\
(\textbf{a}) & (\textbf{b})
\end{tabular}
\caption{(\textbf{a}) Plot of the expectation value of velocity
$\braket{\hat{v}}_{\psi(\tau)}$ as a function of the expectation value of
momentum $p_0$. (\textbf{b}) Plot of the uncertainty in position
$\Delta x_{\psi(\tau)}$ as a function of time $\tau$. Both plots are for a
Gaussian wave packet $\psi(\tau,x)$ with $\Delta p = 1/\sqrt{2}$ and $p_0 = 3$,
and for mass $m = 2$, $\hbar = 1$, $\kappa = 0.1$ (blue plots).}
\label{fig:3}
\end{figure}

In Figure~\ref{fig:2}, plots of the probability density of the state
\eqref{eq:18} in the form of a Gaussian wave packet are presented. From these
plots, we can see that the velocity and spreading of a Gaussian wave packet are
smaller for the $\kappa > 0$ case. In Figure~\ref{fig:3}, a plot of the
expectation value of velocity as a function of the expectation value of momentum
and a plot of the uncertainty in position as a function of time are given. From
these plots, we can see that in the $\kappa > 0$ case, the expectation value of
velocity attains a maximum and approaches $0$ as $p_0 \to \pm\infty$ and that
the spreading of the Gaussian wave packet is slower than in the $\kappa = 0$
case.

\unskip

\subsection{Harmonic Oscillator}
\label{subsec:4.3}
In this section, we will investigate how the minimal time scale influences the
time evolution of a harmonic oscillator initially in a coherent state.
The Hilbert space of the system is $\mathcal{H}_S = L^2(\mathbb{R},\dd{x})$ and
the Hamiltonian of the system is
\begin{equation}
\hat{H}_S = \frac{1}{2m} \hat{p}^2 + \frac{1}{2}m\omega^2 \hat{x}^2,
\end{equation}
where $m$ is the mass of a particle and $\omega$ is a frequency of oscillations.
Let us assume that the system is initially in a coherent state $\ket{\alpha}_S$.
A coherent state $\ket{\alpha}_S$ is defined to be the eigenstate of the
annihilation operator
\begin{equation}
\hat{a} = \sqrt{\frac{m\omega}{2\hbar}} \hat{x} + \frac{i}{\sqrt{2m\omega\hbar}} \hat{p}
\end{equation}
with a corresponding eigenvalue $\alpha \in \mathbb{C}$:
\begin{equation}
\hat{a} \ket{\alpha}_S = \alpha \ket{\alpha}_S.
\end{equation}

The expectation values $x_0,p_0$ of the position and momentum operators in the
coherent state $\ket{\alpha}_S$ are equal
\begin{equation}
x_0 = {}_S\bra{\alpha} \hat{x} \ket{\alpha}_S = \sqrt{\frac{2\hbar}{m\omega}} \re(\alpha), \quad
p_0 = {}_S\bra{\alpha} \hat{p} \ket{\alpha}_S = \sqrt{2m\omega\hbar} \im(\alpha).
\end{equation}

Therefore, $\alpha$ is expressed by the expectation values $x_0,p_0$ of position
and momentum operators in the following way:
\begin{equation}
\alpha = \sqrt{\frac{m\omega}{2\hbar}} x_0 + \frac{i}{\sqrt{2m\omega\hbar}} p_0.
\end{equation}

In the case without a minimal time scale ($\kappa = 0$), the coherent state
$\ket{\alpha}_S$ will evolve in time according to the formula
\begin{equation}
\ket{\alpha(t)}_S = e^{-it\hat{H}_S/\hbar} \ket{\alpha}_S.
\end{equation}

Expanding $\ket{\alpha}_S$ in the basis of eigenstates $\ket{n}_S$ of the
Hamiltonian $\hat{H}_S$, we obtain
\begin{equation}
\ket{\alpha(t)}_S = e^{-it\hat{H}_S/\hbar} \sum_{n=0}^\infty \ket{n}_S {}_S\braket{n|\alpha}_S
= \sum_{n=0}^\infty {}_S\braket{n|\alpha}_S e^{-itE_n/\hbar} \ket{n}_S.
\label{eq:20}
\end{equation}

Using the fact that
\begin{equation}
{}_S\braket{n|\alpha}_S = e^{-\frac{1}{2}\abs{\alpha}^2} \frac{\alpha^n}{\sqrt{n!}}
\end{equation}
and the energy eigenvalues $E_n = \hbar\omega(n + \tfrac{1}{2})$, we can write
\eqref{eq:20} in the form
\begin{linenomath}
\begin{align}
\ket{\alpha(t)}_S & = e^{-\frac{1}{2}\abs{\alpha}^2} \sum_{n=0}^\infty \frac{\alpha^n}{\sqrt{n!}} e^{-i(n + \frac{1}{2})\omega t} \ket{n}_S
= e^{-i\omega t/2} e^{-\frac{1}{2}\abs{\alpha}^2} \sum_{n=0}^\infty \frac{(\alpha e^{-i\omega t})^n}{\sqrt{n!}} \ket{n}_S \nonumber \\
& = e^{-i\omega t/2} \ket{\alpha e^{-i\omega t}}_S.
\label{eq:21}
\end{align}
\end{linenomath}

Therefore, time-evolved coherent states are also coherent with a phase shift.
The position representation of the coherent state $\ket{\alpha}_S$ is equal to
\begin{equation}
\psi_\alpha(x) = {}_S\braket{x|\alpha}_S
= \left(\frac{m\omega}{\pi\hbar}\right)^{1/4} e^{-\frac{m\omega}{2\hbar}(x - x_0)^2} e^{\frac{i}{\hbar}(x - \frac{1}{2}x_0)p_0}.
\label{eq:22}
\end{equation}

From \eqref{eq:21} and \eqref{eq:22}, we obtain
\begin{equation}
\psi_\alpha(t,x) = {}_S\braket{x|\alpha(t)}_S
= \left(\frac{m\omega}{\pi\hbar}\right)^{1/4} e^{-\frac{m\omega}{2\hbar}(x - x_0(t))^2} e^{\frac{i}{\hbar}(x - \frac{1}{2}x_0(t))p_0(t)} e^{-i\omega t/2},
\end{equation}
where $x_0(t)$ and $p_0(t)$ are expectation values of position and momentum
operators in the time-evolved coherent state $\ket{\alpha(t)}_S$ given by
\begin{equation}
\begin{aligned}
x_0(t) & = \sqrt{\frac{2\hbar}{m\omega}} \abs{\alpha}\cos(\omega t - \phi), \\
p_0(t) & = -\sqrt{2m\omega\hbar} \abs{\alpha}\sin(\omega t - \phi),
\end{aligned}
\end{equation}
where $\alpha = \abs{\alpha}e^{i\phi}$.

In the case with a minimal time scale ($\kappa > 0$), the coherent state
$\ket{\alpha}_S$ will evolve in time according to the formula
\begin{equation}
\ket{\alpha(\tau)}_S = \exp\left(-i\tau\frac{1}{\sqrt\kappa}\arctan(\sqrt\kappa \hat{H}_S/\hbar)\right) \ket{\alpha}_S.
\end{equation}

Performing similar calculations as before, we obtain
\begin{equation}
\ket{\alpha(\tau)}_S = e^{-\frac{1}{2}\abs{\alpha}^2} \sum_{n=0}^\infty \frac{\alpha^n}{\sqrt{n!}} e^{-i\tau\frac{1}{\sqrt\kappa}\arctan(\sqrt\kappa\omega(n + \frac{1}{2}))} \ket{n}_S.
\end{equation}

From this, we can see that the state $\ket{\alpha(\tau)}_S$ is no longer
coherent for $\tau \neq 0$. In the position representation, the eigenstates
$\ket{n}_S$ of the Hamiltonian take the form
\begin{equation}
\psi_n(x) = {}_S\braket{x|n}_S = \frac{1}{\sqrt{2^n n!}} \left(\frac{m\omega}{\pi\hbar}\right)^{1/4} e^{-\frac{m\omega x^2}{2\hbar}} H_n\left(\sqrt{\frac{m\omega}{\hbar}}x\right),
\end{equation}
where $H_n$ are Hermite polynomials. Hence, the time-evolved coherent state
$\ket{\alpha(\tau)}_S$ takes the form
\begin{linenomath}
\begin{align}
\psi_\alpha(\tau,x) & = {}_S\braket{x|\alpha(\tau)}_S \nonumber \\
& = \left(\frac{m\omega}{\pi\hbar}\right)^{1/4} e^{-\frac{m\omega x^2}{2\hbar}}
    e^{-\frac{1}{2}\abs{\alpha}^2} \sum_{n=0}^\infty \frac{1}{n!}
    \left(\frac{\alpha}{\sqrt{2}}\right)^n H_n\left(\sqrt{\frac{m\omega}{\hbar}}x\right)
    e^{-i\tau\frac{1}{\sqrt\kappa}\arctan(\sqrt\kappa\omega(n + \frac{1}{2}))}.
\label{eq:23}
\end{align}
\end{linenomath}

\begin{figure}
\begin{tabular}{cc}
\includegraphics{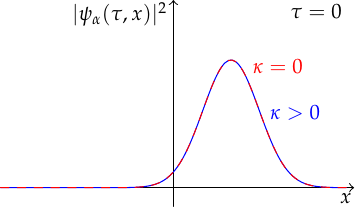}
%\begin{asy}
%usepackage("mathpazo");
%import graph;
%size(170,100,IgnoreAspect);
%defaultpen(fontsize(10pt));

%real t = 0;
%real kappa = 0.01;
%real m = 1;
%real omega = 2*pi/3;
%real x0 = 1;
%real p0 = 0;

%real H(int n, real x) {
%  real sum = 0;
%  for(int m = 0; m <= floor(n/2); ++m)
%    sum += (-1)^m * (2*x)^(n-2*m) / (factorial(m)*factorial(n-2*m));
%  return sum;
%}

%pair psi(real t, real x) {
%  pair sum = (0,0);
%  pair alpha = (sqrt(m*omega/2)*x0,1/sqrt(2*m*omega)*p0);
%  for(int n = 0; n <= 20; ++n)
%    sum += (alpha/sqrt(2))^n * H(n,sqrt(m*omega)*x) * expi(-t*atan(sqrt(kappa)*(n+1/2)*omega)/sqrt(kappa));
%  return sqrt(sqrt(m*omega/pi)) * exp(-m*omega*x^2/2) * exp(-abs(alpha)^2/2) * sum;
%}

%real phi2(real t, real x) {
%  pair alpha = (sqrt(m*omega/2)*x0,1/sqrt(2*m*omega)*p0);
%  return sqrt(m*omega/pi) * exp(-m*omega*(x - sqrt(2/(m*omega))*abs(alpha)*cos(atan(alpha.y/alpha.x) - omega*t))^2);
%}

%draw(graph(new real(real x) { return abs(psi(t,x))^2; },-3,3,operator ..),blue);
%draw(graph(new real(real x) { return phi2(t,x); },-3,3,operator ..),red+dashed);
%xaxis("$x$",Arrow(TeXHead));
%yaxis("$|\psi_\alpha(\tau,x)|^2$",ymax=1.2,Arrow(TeXHead),autorotate=false);
%label("$\tau = 0$",(3,1.2),SW);
%label("$\kappa = 0$",(1.3,0.7),NE,red);
%label("$\kappa > 0$",(1.6,0.4),NE,blue);
%\end{asy}
&
\includegraphics{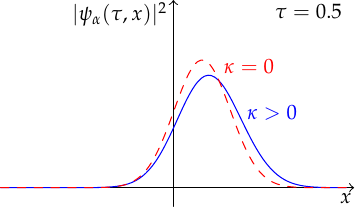}
%\begin{asy}
%usepackage("mathpazo");
%import graph;
%size(170,100,IgnoreAspect);
%defaultpen(fontsize(10pt));

%real t = 0.5;
%real kappa = 0.01;
%real m = 1;
%real omega = 2*pi/3;
%real x0 = 1;
%real p0 = 0;

%real H(int n, real x) {
%  real sum = 0;
%  for(int m = 0; m <= floor(n/2); ++m)
%    sum += (-1)^m * (2*x)^(n-2*m) / (factorial(m)*factorial(n-2*m));
%  return sum;
%}

%pair psi(real t, real x) {
%  pair sum = (0,0);
%  pair alpha = (sqrt(m*omega/2)*x0,1/sqrt(2*m*omega)*p0);
%  for(int n = 0; n <= 20; ++n)
%    sum += (alpha/sqrt(2))^n * H(n,sqrt(m*omega)*x) * expi(-t*atan(sqrt(kappa)*(n+1/2)*omega)/sqrt(kappa));
%  return sqrt(sqrt(m*omega/pi)) * exp(-m*omega*x^2/2) * exp(-abs(alpha)^2/2) * sum;
%}

%real phi2(real t, real x) {
%  pair alpha = (sqrt(m*omega/2)*x0,1/sqrt(2*m*omega)*p0);
%  return sqrt(m*omega/pi) * exp(-m*omega*(x - sqrt(2/(m*omega))*abs(alpha)*cos(atan(alpha.y/alpha.x) - omega*t))^2);
%}

%draw(graph(new real(real x) { return abs(psi(t,x))^2; },-3,3,operator ..),blue);
%draw(graph(new real(real x) { return phi2(t,x); },-3,3,operator ..),red+dashed);
%xaxis("$x$",Arrow(TeXHead));
%yaxis("$|\psi_\alpha(\tau,x)|^2$",ymax=1.2,Arrow(TeXHead),autorotate=false);
%label("$\tau = 0.5$",(3,1.2),SW);
%label("$\kappa = 0$",(0.8,0.7),NE,red);
%label("$\kappa > 0$",(1.2,0.4),NE,blue);
%\end{asy}
\\ (\textbf{a}) & (\textbf{b}) \\ \\
\includegraphics{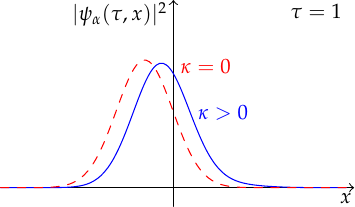}
%\begin{asy}
%usepackage("mathpazo");
%import graph;
%size(170,100,IgnoreAspect);
%defaultpen(fontsize(10pt));

%real t = 1;
%real kappa = 0.01;
%real m = 1;
%real omega = 2*pi/3;
%real x0 = 1;
%real p0 = 0;

%real H(int n, real x) {
%  real sum = 0;
%  for(int m = 0; m <= floor(n/2); ++m)
%    sum += (-1)^m * (2*x)^(n-2*m) / (factorial(m)*factorial(n-2*m));
%  return sum;
%}

%pair psi(real t, real x) {
%  pair sum = (0,0);
%  pair alpha = (sqrt(m*omega/2)*x0,1/sqrt(2*m*omega)*p0);
%  for(int n = 0; n <= 20; ++n)
%    sum += (alpha/sqrt(2))^n * H(n,sqrt(m*omega)*x) * expi(-t*atan(sqrt(kappa)*(n+1/2)*omega)/sqrt(kappa));
%  return sqrt(sqrt(m*omega/pi)) * exp(-m*omega*x^2/2) * exp(-abs(alpha)^2/2) * sum;
%}

%real phi2(real t, real x) {
%  pair alpha = (sqrt(m*omega/2)*x0,1/sqrt(2*m*omega)*p0);
%  return sqrt(m*omega/pi) * exp(-m*omega*(x - sqrt(2/(m*omega))*abs(alpha)*cos(atan(alpha.y/alpha.x) - omega*t))^2);
%}

%draw(graph(new real(real x) { return abs(psi(t,x))^2; },-3,3,operator ..),blue);
%draw(graph(new real(real x) { return phi2(t,x); },-3,3,operator ..),red+dashed);
%xaxis("$x$",Arrow(TeXHead));
%yaxis("$|\psi_\alpha(\tau,x)|^2$",ymax=1.2,Arrow(TeXHead),autorotate=false);
%label("$\tau = 1$",(3,1.2),SW);
%label("$\kappa = 0$",(0.05,0.7),NE,red);
%label("$\kappa > 0$",(0.35,0.4),NE,blue);
%\end{asy}
&
\includegraphics{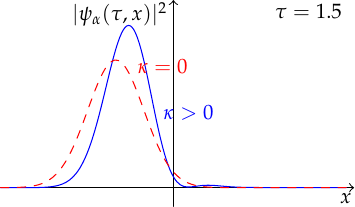}
%\begin{asy}
%usepackage("mathpazo");
%import graph;
%size(170,100,IgnoreAspect);
%defaultpen(fontsize(10pt));

%real t = 1.5;
%real kappa = 0.01;
%real m = 1;
%real omega = 2*pi/3;
%real x0 = 1;
%real p0 = 0;

%real H(int n, real x) {
%  real sum = 0;
%  for(int m = 0; m <= floor(n/2); ++m)
%    sum += (-1)^m * (2*x)^(n-2*m) / (factorial(m)*factorial(n-2*m));
%  return sum;
%}

%pair psi(real t, real x) {
%  pair sum = (0,0);
%  pair alpha = (sqrt(m*omega/2)*x0,1/sqrt(2*m*omega)*p0);
%  for(int n = 0; n <= 20; ++n)
%    sum += (alpha/sqrt(2))^n * H(n,sqrt(m*omega)*x) * expi(-t*atan(sqrt(kappa)*(n+1/2)*omega)/sqrt(kappa));
%  return sqrt(sqrt(m*omega/pi)) * exp(-m*omega*x^2/2) * exp(-abs(alpha)^2/2) * sum;
%}

%real phi2(real t, real x) {
%  pair alpha = (sqrt(m*omega/2)*x0,1/sqrt(2*m*omega)*p0);
%  return sqrt(m*omega/pi) * exp(-m*omega*(x - sqrt(2/(m*omega))*abs(alpha)*cos(atan(alpha.y/alpha.x) - omega*t))^2);
%}

%draw(graph(new real(real x) { return abs(psi(t,x))^2; },-3,3,operator ..),blue);
%draw(graph(new real(real x) { return phi2(t,x); },-3,3,operator ..),red+dashed);
%xaxis("$x$",Arrow(TeXHead));
%yaxis("$|\psi_\alpha(\tau,x)|^2$",ymax=1.2,Arrow(TeXHead),autorotate=false);
%label("$\tau = 1.5$",(3,1.2),SW);
%label("$\kappa = 0$",(-0.7,0.7),NE,red);
%label("$\kappa > 0$",(-0.25,0.4),NE,blue);
%\end{asy}
\\ (\textbf{c}) & (\textbf{d}) \\ \\
\includegraphics{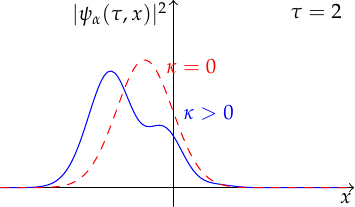}
%\begin{asy}
%usepackage("mathpazo");
%import graph;
%size(170,100,IgnoreAspect);
%defaultpen(fontsize(10pt));

%real t = 2;
%real kappa = 0.01;
%real m = 1;
%real omega = 2*pi/3;
%real x0 = 1;
%real p0 = 0;

%real H(int n, real x) {
%  real sum = 0;
%  for(int m = 0; m <= floor(n/2); ++m)
%    sum += (-1)^m * (2*x)^(n-2*m) / (factorial(m)*factorial(n-2*m));
%  return sum;
%}

%pair psi(real t, real x) {
%  pair sum = (0,0);
%  pair alpha = (sqrt(m*omega/2)*x0,1/sqrt(2*m*omega)*p0);
%  for(int n = 0; n <= 20; ++n)
%    sum += (alpha/sqrt(2))^n * H(n,sqrt(m*omega)*x) * expi(-t*atan(sqrt(kappa)*(n+1/2)*omega)/sqrt(kappa));
%  return sqrt(sqrt(m*omega/pi)) * exp(-m*omega*x^2/2) * exp(-abs(alpha)^2/2) * sum;
%}

%real phi2(real t, real x) {
%  pair alpha = (sqrt(m*omega/2)*x0,1/sqrt(2*m*omega)*p0);
%  return sqrt(m*omega/pi) * exp(-m*omega*(x - sqrt(2/(m*omega))*abs(alpha)*cos(atan(alpha.y/alpha.x) - omega*t))^2);
%}

%draw(graph(new real(real x) { return abs(psi(t,x))^2; },-3,3,operator ..),blue);
%draw(graph(new real(real x) { return phi2(t,x); },-3,3,operator ..),red+dashed);
%xaxis("$x$",Arrow(TeXHead));
%yaxis("$|\psi_\alpha(\tau,x)|^2$",ymax=1.2,Arrow(TeXHead),autorotate=false);
%label("$\tau = 2$",(3,1.2),SW);
%label("$\kappa = 0$",(-0.2,0.7),NE,red);
%label("$\kappa > 0$",(0.1,0.4),NE,blue);
%\end{asy}
&
\includegraphics{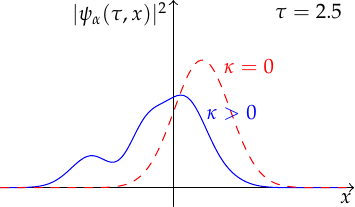}
%\begin{asy}
%usepackage("mathpazo");
%import graph;
%size(170,100,IgnoreAspect);
%defaultpen(fontsize(10pt));

%real t = 2.5;
%real kappa = 0.01;
%real m = 1;
%real omega = 2*pi/3;
%real x0 = 1;
%real p0 = 0;

%real H(int n, real x) {
%  real sum = 0;
%  for(int m = 0; m <= floor(n/2); ++m)
%    sum += (-1)^m * (2*x)^(n-2*m) / (factorial(m)*factorial(n-2*m));
%  return sum;
%}

%pair psi(real t, real x) {
%  pair sum = (0,0);
% pair alpha = (sqrt(m*omega/2)*x0,1/sqrt(2*m*omega)*p0);
%  for(int n = 0; n <= 20; ++n)
%    sum += (alpha/sqrt(2))^n * H(n,sqrt(m*omega)*x) * expi(-t*atan(sqrt(kappa)*(n+1/2)*omega)/sqrt(kappa));
%  return sqrt(sqrt(m*omega/pi)) * exp(-m*omega*x^2/2) * exp(-abs(alpha)^2/2) * sum;
%}

%real phi2(real t, real x) {
%  pair alpha = (sqrt(m*omega/2)*x0,1/sqrt(2*m*omega)*p0);
%  return sqrt(m*omega/pi) * exp(-m*omega*(x - sqrt(2/(m*omega))*abs(alpha)*cos(atan(alpha.y/alpha.x) - omega*t))^2);
%}

%draw(graph(new real(real x) { return abs(psi(t,x))^2; },-3,3,operator ..),blue);
%draw(graph(new real(real x) { return phi2(t,x); },-3,3,operator ..),red+dashed);
%xaxis("$x$",Arrow(TeXHead));
%yaxis("$|\psi_\alpha(\tau,x)|^2$",ymax=1.2,Arrow(TeXHead),autorotate=false);
%label("$\tau = 2.5$",(3,1.2),SW);
%label("$\kappa = 0$",(0.8,0.7),NE,red);
%label("$\kappa > 0$",(0.5,0.4),NE,blue);
%\end{asy}
\\ (\textbf{e}) & (\textbf{f}) \\ \\
\includegraphics{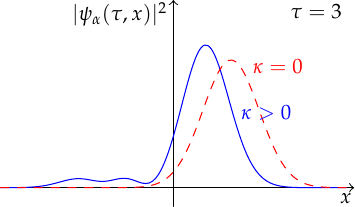}
%\begin{asy}
%usepackage("mathpazo");
%import graph;
%size(170,100,IgnoreAspect);
%defaultpen(fontsize(10pt));

%real t = 3;
%real kappa = 0.01;
%real m = 1;
%real omega = 2*pi/3;
%real x0 = 1;
%real p0 = 0;

%real H(int n, real x) {
%  real sum = 0;
%  for(int m = 0; m <= floor(n/2); ++m)
%    sum += (-1)^m * (2*x)^(n-2*m) / (factorial(m)*factorial(n-2*m));
%  return sum;
%}

%pair psi(real t, real x) {
%  pair sum = (0,0);
%  pair alpha = (sqrt(m*omega/2)*x0,1/sqrt(2*m*omega)*p0);
%  for(int n = 0; n <= 20; ++n)
%    sum += (alpha/sqrt(2))^n * H(n,sqrt(m*omega)*x) * expi(-t*atan(sqrt(kappa)*(n+1/2)*omega)/sqrt(kappa));
%  return sqrt(sqrt(m*omega/pi)) * exp(-m*omega*x^2/2) * exp(-abs(alpha)^2/2) * sum;
%}

%real phi2(real t, real x) {
%  pair alpha = (sqrt(m*omega/2)*x0,1/sqrt(2*m*omega)*p0);
%  return sqrt(m*omega/pi) * exp(-m*omega*(x - sqrt(2/(m*omega))*abs(alpha)*cos(atan(alpha.y/alpha.x) - omega*t))^2);
%}

%draw(graph(new real(real x) { return abs(psi(t,x))^2; },-3,3,operator ..),blue);
%draw(graph(new real(real x) { return phi2(t,x); },-3,3,operator ..),red+dashed);
%xaxis("$x$",Arrow(TeXHead));
%yaxis("$|\psi_\alpha(\tau,x)|^2$",ymax=1.2,Arrow(TeXHead),autorotate=false);
%label("$\tau = 3$",(3,1.2),SW);
%label("$\kappa = 0$",(1.3,0.7),NE,red);
%label("$\kappa > 0$",(1.1,0.4),NE,blue);
%\end{asy}
&
\includegraphics{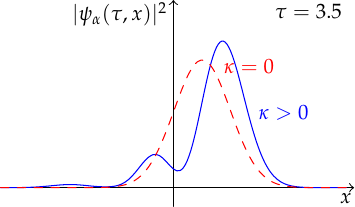}
%\begin{asy}
%usepackage("mathpazo");
%import graph;
%size(170,100,IgnoreAspect);
%defaultpen(fontsize(10pt));

%real t = 3.5;
%real kappa = 0.01;
%real m = 1;
%real omega = 2*pi/3;
%real x0 = 1;
%real p0 = 0;

%real H(int n, real x) {
%  real sum = 0;
%  for(int m = 0; m <= floor(n/2); ++m)
%    sum += (-1)^m * (2*x)^(n-2*m) / (factorial(m)*factorial(n-2*m));
%  return sum;
%}

%pair psi(real t, real x) {
%  pair sum = (0,0);
%  pair alpha = (sqrt(m*omega/2)*x0,1/sqrt(2*m*omega)*p0);
%  for(int n = 0; n <= 20; ++n)
%    sum += (alpha/sqrt(2))^n * H(n,sqrt(m*omega)*x) * expi(-t*atan(sqrt(kappa)*(n+1/2)*omega)/sqrt(kappa));
%  return sqrt(sqrt(m*omega/pi)) * exp(-m*omega*x^2/2) * exp(-abs(alpha)^2/2) * sum;
%}

%real phi2(real t, real x) {
%  pair alpha = (sqrt(m*omega/2)*x0,1/sqrt(2*m*omega)*p0);
%  return sqrt(m*omega/pi) * exp(-m*omega*(x - sqrt(2/(m*omega))*abs(alpha)*cos(atan(alpha.y/alpha.x) - omega*t))^2);
%}

%draw(graph(new real(real x) { return abs(psi(t,x))^2; },-3,3,operator ..),blue);
%draw(graph(new real(real x) { return phi2(t,x); },-3,3,operator ..),red+dashed);
%xaxis("$x$",Arrow(TeXHead));
%yaxis("$|\psi_\alpha(\tau,x)|^2$",ymax=1.2,Arrow(TeXHead),autorotate=false);
%label("$\tau = 3.5$",(3,1.2),SW);
%label("$\kappa = 0$",(0.8,0.7),NE,red);
%label("$\kappa > 0$",(1.4,0.4),NE,blue);
%\end{asy}
\\ (\textbf{g}) & (\textbf{h})
\end{tabular}
\caption{(\textbf{a}--\textbf{h}) Plots of the probability density of the
time-evolved coherent state $\psi_\alpha(\tau,x)$ with $x_0 = 1$ and $p_0 = 0$
for particular instances of time $\tau$. Mass $m = 1$, frequency
$\omega = 2\pi/3$, $\hbar = 1$, $\kappa = 0.01$ (blue~plots).}
\label{fig:4}
\end{figure}

In Figure~\ref{fig:4}, plots of the probability density of the state
\eqref{eq:23} for particular values of time $\tau$ are presented. These plots
illustrate how the coherence of the state is destroyed during time evolution as
a result of the existence of the fundamental limit for the precision with which
we can measure time.

\subsection{System of Two Coupled Spins-$1/2$ in an External Magnetic Field}
\label{subsec:4.4}
In this section, we will investigate a system composed of two spins-$1/2$
interacting with each other and with an external constant magnetic field. The
interaction between spins will cause the system initially in a non-entangled
state to entangle over time. We will explore how the minimal time scale
influences the entanglement.

The Hilbert space of the system is $\mathcal{H}_S = \mathcal{H}_1 \otimes
\mathcal{H}_2 = \mathbb{C}^2 \otimes \mathbb{C}^2 = \mathbb{C}^4$, where
$\mathcal{H}_1 = \mathbb{C}^2$ and $\mathcal{H}_2 = \mathbb{C}^2$ are Hilbert
spaces of individual spins. We will take the following Hamiltonian of the
system:
\begin{linenomath}
\begin{align}
\hat{H}_S & = \hat{H}_0 + \hat{H}_1
= -\gamma B_0 (\hat{S}_z \otimes \hat{1} + \hat{1} \otimes \hat{S}_z)
    + \lambda(\hat{S}_x \otimes \hat{S}_x + \hat{S}_y \otimes \hat{S}_y) \nonumber \\
& = \frac{\hbar\omega_0}{2} (\sigma_z \otimes I + I \otimes \sigma_z)
    + \frac{\hbar^2 \lambda}{4} (\sigma_x \otimes \sigma_x + \sigma_y \otimes \sigma_y),
\end{align}
\end{linenomath}
where $\gamma$ is the gyromagnetic ratio; $B_0$ is the strength of the magnetic
field, which we assume is directed along the \nbr{z}axis; $\lambda$ is the
strength of the interaction between spins; $\omega_0 = -\gamma B_0$ is the
frequency of precession of the spins; and $\hat{S}_x$, $\hat{S}_y$, $\hat{S}_z$
are components of the spin operator expressed by the Pauli matrices
\begin{equation}
\sigma_x = \begin{bmatrix}
0 & 1 \\
1 & 0
\end{bmatrix}, \quad
\sigma_y = \begin{bmatrix}
0 & -i \\
i & 0
\end{bmatrix}, \quad
\sigma_z = \begin{bmatrix}
1 & 0 \\
0 & -1
\end{bmatrix}.
\end{equation}

The operator $\hat{H}_0$ describes non-interacting spins precessing around the
\nbr{z}axis with the frequency $\omega_0$. The operator $\hat{H}_1$ describes
the interaction between spins.

First, let us notice that $\hat{H}_0 \hat{H}_1 = \hat{H}_1 \hat{H}_0 = 0$.
Because of this, $\hat{H}_S^n = \hat{H}_0^n + \hat{H}_1^n$ for
$n \in \mathbb{N}$ and
\begin{equation}
\frac{1}{\sqrt\kappa} \arctan(\sqrt\kappa \hat{H}_S/\hbar) =
    \frac{1}{\sqrt\kappa} \arctan(\sqrt\kappa \hat{H}_0/\hbar)
    + \frac{1}{\sqrt\kappa} \arctan(\sqrt\kappa \hat{H}_1/\hbar).
\end{equation}

Thus
\begin{linenomath}
\begin{multline}
\exp\left(-i\tau\frac{1}{\sqrt\kappa} \arctan(\sqrt\kappa \hat{H}_S/\hbar)\right) =
    \exp\left(-i\tau\frac{1}{\sqrt\kappa} \arctan(\sqrt\kappa \hat{H}_1/\hbar)\right) \\
{} \times \exp\left(-i\tau\frac{1}{\sqrt\kappa} \arctan(\sqrt\kappa \hat{H}_0/\hbar)\right).
\end{multline}
\end{linenomath}

Next, observe that for $n \in \mathbb{N}$
\begin{equation}
\begin{aligned}
\hat{H}_0^{2n+1} & = \frac{1}{2} (\hbar\omega_0)^{2n+1} (\sigma_z \otimes I + I \otimes \sigma_z), \\
\hat{H}_1^{2n+1} & = \frac{1}{2} \left(\frac{\hbar^2 \lambda}{2}\right)^{2n+1} (\sigma_x \otimes \sigma_x + \sigma_y \otimes \sigma_y).
\end{aligned}
\end{equation}

From this,
\begin{equation}
\begin{aligned}
\frac{1}{\sqrt\kappa} \arctan(\sqrt\kappa \hat{H}_0/\hbar) & = \frac{1}{2\sqrt\kappa} \arctan(\sqrt\kappa\omega_0) (\sigma_z \otimes I + I \otimes \sigma_z), \\
\frac{1}{\sqrt\kappa} \arctan(\sqrt\kappa \hat{H}_1/\hbar) & = \frac{1}{2\sqrt\kappa} \arctan\left(\frac{\hbar\sqrt\kappa}{2}\lambda\right) (\sigma_x \otimes \sigma_x + \sigma_y \otimes \sigma_y).
\end{aligned}
\end{equation}

Ultimately, the time evolution of the system initially in a non-entangled
state\linebreak $\ket{\psi}_S = \ket{\psi_1} \otimes \ket{\psi_2}$ is given by
the formula
\begin{equation}
\ket{\psi(\tau)}_S = e^{-i\tau\hbar\lambda_\kappa (\sigma_x \otimes \sigma_x + \sigma_y \otimes \sigma_y)/4}
    e^{-i\tau\omega_\kappa (\sigma_z \otimes I + I \otimes \sigma_z)/2} \ket{\psi_1} \otimes \ket{\psi_2},
\label{eq:26}
\end{equation}
where $\omega_\kappa = \frac{1}{\sqrt\kappa} \arctan(\sqrt\kappa\omega_0)$ and
$\lambda_\kappa = \frac{2}{\hbar\sqrt\kappa} \arctan\left(\frac{\hbar\sqrt\kappa}{2}\lambda\right)$.

Let us assume that the initial states $\ket{\psi_1}$ and $\ket{\psi_2}$ are the
same and equal
\begin{equation}
\ket{\psi_1} = \ket{\psi_2} = \ket{\theta,\varphi} = \cos(\theta/2) e^{-i\varphi/2} \ket{\uparrow} + \sin(\theta/2) e^{i\varphi/2} \ket{\downarrow},
\end{equation}
where $\theta,\varphi \in \mathbb{R}$ and $\ket{\uparrow}$, $\ket{\downarrow}$
are spin-up and spin-down eigenstates of the spin operator $\hat{S}_z$.
The action of the operator $e^{-i\tau\omega_\kappa (\sigma_z \otimes I + I \otimes \sigma_z)/2}$
on $\ket{\psi_1} \otimes \ket{\psi_2}$ will result in each of the states
$\ket{\psi_1}$ and $\ket{\psi_2}$ evolving in time according to the formula
\eqref{eq:24}, so only the angle $\varphi$ will change in both of these states:
\begin{equation}
e^{-i\tau\omega_\kappa (\sigma_z \otimes I + I \otimes \sigma_z)/2} \ket{\psi_1} \otimes \ket{\psi_2} = \ket{\theta,\varphi + \omega_\kappa \tau} \otimes \ket{\theta,\varphi + \omega_\kappa \tau}.
\end{equation}

We can calculate that
\begin{linenomath}
\begin{multline}
e^{-i\tau\hbar\lambda_\kappa (\sigma_x \otimes \sigma_x + \sigma_y \otimes \sigma_y)/4} =
    \frac{1}{2} \biggl( I \otimes I + \sigma_z \otimes \sigma_z
    + \cos\left(\frac{\hbar\lambda_\kappa \tau}{2}\right) (I \otimes I - \sigma_z \otimes \sigma_z) \\
    {} - i\sin\left(\frac{\hbar\lambda_\kappa \tau}{2}\right) (\sigma_x \otimes \sigma_x + \sigma_y \otimes \sigma_y) \biggr).
\end{multline}
\end{linenomath}

From this, we obtain the following expression for the time evolution of the
state $\ket{\psi}_S = \ket{\psi_1} \otimes \ket{\psi_2}$
\begin{linenomath}
\begin{multline}
\ket{\psi(\tau)}_S = \cos^2\frac{\theta}{2} e^{-i(\varphi + \omega_\kappa \tau)} \ket{\uparrow} \otimes \ket{\uparrow}
    + \sin^2\frac{\theta}{2} e^{i(\varphi + \omega_\kappa \tau)} \ket{\downarrow} \otimes \ket{\downarrow} \\
{} + \frac{1}{2} \sin\theta e^{-i\hbar\lambda_\kappa\tau/2} (\ket{\uparrow} \otimes \ket{\downarrow} + \ket{\downarrow} \otimes \ket{\uparrow}).
\label{eq:25}
\end{multline}
\end{linenomath}

A useful measure of the degree of quantum entanglement between two subsystems is
the entropy of entanglement $S$. It is defined as the Von Neumann entropy of the
reduced density matrix for any of the subsystems:
\begin{equation}
S = -k_B \Tr(\hat{\rho}_1 \ln\hat{\rho}_1) = -k_B \Tr(\hat{\rho}_2 \ln\hat{\rho}_2),
\end{equation}
where $k_B$ is the Boltzmann's constant, $\hat{\rho}_1 = \Tr_2(\hat{\rho})$ and
$\hat{\rho}_2 = \Tr_1(\hat{\rho})$ are the reduced density matrices of each
subsystem, and $\hat{\rho}$ is the density matrix of the whole system. If the
subsystems are not entangled, then $S = 0$; otherwise, $S > 0$, and the bigger
the entropy $S$ is, the more entangled the subsystems are.

The density matrix of the system of two spins is equal $\hat{\rho} =
\ket{\psi(\tau)}\bra{\psi(\tau)}_S$, where $\ket{\psi(\tau)}_S$ is given by
\eqref{eq:25}. By taking the partial trace $\Tr_2$ of $\hat{\rho}$ over the
Hilbert space of the second subsystem, we receive the reduced density matrix
$\hat{\rho}_1$ of the first subsystem. It is given by the equation
\begin{linenomath}
\begin{multline}
\hat{\rho}_1 = \cos^2\frac{\theta}{2} \ket{\uparrow}\bra{\uparrow} + \sin^2\frac{\theta}{2} \ket{\downarrow}\bra{\downarrow} \\
    {} + \frac{1}{2}\sin\theta e^{-i(\varphi + \omega_\kappa \tau)} \left(\cos^2\frac{\theta}{2} e^{i\hbar\lambda_\kappa\tau/2}
    + \sin^2\frac{\theta}{2} e^{-i\hbar\lambda_\kappa\tau/2}\right) \ket{\uparrow}\bra{\downarrow} \\
    {} + \frac{1}{2}\sin\theta e^{i(\varphi + \omega_\kappa \tau)} \left(\cos^2\frac{\theta}{2} e^{-i\hbar\lambda_\kappa\tau/2}
    + \sin^2\frac{\theta}{2} e^{i\hbar\lambda_\kappa\tau/2}\right) \ket{\downarrow}\bra{\uparrow}.
\end{multline}
\end{linenomath}

The matrix $\hat{\rho}_1$ has the following eigenvalues
\begin{equation}
p_1 = \frac{1 + \sqrt{1 - \sin^4\theta \sin^2(\hbar\lambda_\kappa\tau/2)}}{2}, \quad
p_2 = \frac{1 - \sqrt{1 - \sin^4\theta \sin^2(\hbar\lambda_\kappa\tau/2)}}{2}.
\end{equation}

Thus, the entropy of entanglement is equal to
\begin{linenomath}
\begin{align}
S & = -k_B \Tr(\hat{\rho}_1 \ln\hat{\rho}_1) = -k_B (p_1 \ln p_1 + p_2 \ln p_2) \nonumber \\
& = -k_B \biggl( \ln\frac{\sin^2\theta \abs{\sin(\hbar\lambda_\kappa\tau/2)}}{4} \nonumber \\
& \quad {} + \sqrt{1 - \sin^4\theta \sin^2(\hbar\lambda_\kappa\tau/2)} \artanh\sqrt{1 - \sin^4\theta \sin^2(\hbar\lambda_\kappa\tau/2)} \biggr).
\end{align}
\end{linenomath}

\begin{figure}[H]
\includegraphics{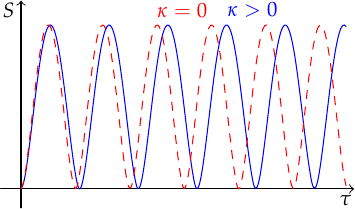}
%\begin{asy}
%usepackage("mathpazo");
%import graph;
%size(170,100,IgnoreAspect);
%defaultpen(fontsize(10pt));

%real kappa = 0.01;
%real theta = pi/4;
%real lambda = 2*atan(10*sqrt(kappa)/2)/sqrt(kappa);

%real S(real t) {
%  real p1 = (1 + sqrt(1 - (sin(theta))^4 * (sin(lambda*t/2))^2))/2;
%  real p2 = (1 - sqrt(1 - (sin(theta))^4 * (sin(lambda*t/2))^2))/2;
%  if (p2 == 0)
%    return 0;
%  return -p1*log(p1) - p2*log(p2);
%}

%draw(graph(S,0,3.75,operator ..),blue);
%lambda = 10;
%draw(graph(S,0,3.75,operator ..),red+dashed);
%xaxis("$\tau$",Arrow(TeXHead));
%yaxis("$S$",Arrow(TeXHead),autorotate=false);
%label("$\kappa = 0$",(2.2,0.25),NW,red);
%label("$\kappa > 0$",(2.33,0.25),NE,blue);
%\end{asy}
\caption{Plots of the entropy of entanglement as a function of time $\tau$ for
$\theta = \pi/4$, $\lambda = 10$, $\hbar = 1$ and $\kappa = 0.01$ (blue plot).}
\label{fig:5}
\end{figure}

From the above formula, we can see that $S$ is a periodic function of $\tau$.
The entanglement of the spins will periodically increase reaching its maximum,
then decrease until the spins become non-entangled. The period of these
oscillations is equal to $\frac{2\pi}{\hbar\lambda_\kappa}$. In the case without
a minimal time scale ($\kappa = 0$), this period is equal to
$\frac{2\pi}{\hbar\lambda}$ and is smaller than in the case with a minimal time
scale ($\kappa > 0$); see Figure~\ref{fig:5}. Therefore, the existence of the
fundamental limit for the precision with which we can measure time influences
the entanglement.

Notice that from \eqref{eq:26} we can see that two non-interacting spins will
precess around the \nbr{z}axis with a smaller frequency than only one spin, as
described in Section~\ref{subsec:4.1}. This result is quite peculiar because it
means that even though the spins are not interacting, they still influence each
other in some way which causes the decrease in the frequency of~precession.

\subsection{System of Three Spins-$1/2$ in an External Magnetic Field}
\label{subsec:4.5}
In this section, we will investigate a system composed of three spins-$1/2$
interacting with an external constant magnetic field. We will show that the
minimal time scale will cause the spins to entangle over time even though the
spins are not interacting with each~other.

The Hilbert space of the system is $\mathcal{H}_S = \mathcal{H}_1 \otimes
\mathcal{H}_2 \otimes \mathcal{H}_3 = \mathbb{C}^2 \otimes \mathbb{C}^2 \otimes
\mathbb{C}^2 = \mathbb{C}^6$, where $\mathcal{H}_1 = \mathbb{C}^2$,
$\mathcal{H}_2 = \mathbb{C}^2$ and $\mathcal{H}_3 = \mathbb{C}^2$ are Hilbert
spaces of individual spins. The Hamiltonian of the system is equal
\begin{linenomath}
\begin{align}
\hat{H}_S & = -\gamma B_0 (\hat{S}_z \otimes \hat{1} \otimes \hat{1} + \hat{1} \otimes \hat{S}_z \otimes \hat{1} + \hat{1} \otimes \hat{1} \otimes \hat{S}_z) \nonumber \\
& = \frac{\hbar\omega_0}{2} (\sigma_z \otimes I \otimes I + I \otimes \sigma_z \otimes I + I \otimes I \otimes \sigma_z),
\end{align}
\end{linenomath}
where $\gamma$ is the gyromagnetic ratio; $B_0$ is the strength of the magnetic
field, which we assume is directed along the \nbr{z}axis;
$\omega_0 = -\gamma B_0$; and $\hat{S}_z$ is the \nbr{z}component of the spin
operator expressed by the Pauli matrix $\sigma_z$.

We can calculate that for $n \in \mathbb{N}$
\begin{linenomath}
\begin{multline}
\hat{H}_S^{2n+1} = \frac{1}{4} \left(\frac{\hbar\omega_0}{2}\right)^{2n+1}
    \Bigl( (3^{2n+1} + 1)(\sigma_z \otimes I \otimes I + I \otimes \sigma_z \otimes I + I \otimes I \otimes \sigma_z) \\
    {} + (3^{2n+1} - 3) \sigma_z \otimes \sigma_z \otimes \sigma_z \Bigr)
\end{multline}
\end{linenomath}
for which we obtain
\begin{linenomath}
\begin{align}
\hat{H}_{\mathrm{eff}} & = \frac{\hbar}{\sqrt\kappa}\arctan(\sqrt\kappa \hat{H}_S/\hbar) \nonumber \\
& = \frac{\hbar\omega_\kappa}{2}(\sigma_z \otimes I \otimes I + I \otimes \sigma_z \otimes I + I \otimes I \otimes \sigma_z)
    + \hbar\lambda_\kappa \sigma_z \otimes \sigma_z \otimes \sigma_z,
\end{align}
\end{linenomath}
where
\begin{equation}
\begin{aligned}
\omega_\kappa & = \frac{1}{4\sqrt\kappa}\left(\arctan\left(\frac{3\sqrt\kappa\omega_0}{2}\right) + \arctan\left(\frac{\sqrt\kappa\omega_0}{2}\right)\right), \\
\lambda_\kappa & = \frac{1}{4\sqrt\kappa}\left(\arctan\left(\frac{3\sqrt\kappa\omega_0}{2}\right) - 3\arctan\left(\frac{\sqrt\kappa\omega_0}{2}\right)\right).
\end{aligned}
\end{equation}

We can see that the effective Hamiltonian $\hat{H}_{\mathrm{eff}}$ describing
the time evolution of the system contains a term
$\hbar\lambda_\kappa \sigma_z \otimes \sigma_z \otimes \sigma_z$ which will
cause the entanglement of the system over time. Taking the system initially in
a state
\begin{equation}
\ket{\psi}_S = \ket{\theta_1,\varphi_1} \otimes \ket{\theta_2,\varphi_2} \otimes \ket{\theta_3,\varphi_3}
\end{equation}
and noticing that
\begin{equation}
e^{-i\tau\lambda_\kappa \sigma_z \otimes \sigma_z \otimes \sigma_z} =
    \cos(\lambda_\kappa \tau) I \otimes I \otimes I - i\sin(\lambda_\kappa \tau) \sigma_z \otimes \sigma_z \otimes \sigma_z
\end{equation}
we obtain the following expression for the time evolution of the state
$\ket{\psi}_S$:
\begin{linenomath}
\begin{multline}
\ket{\psi(\tau)}_S = \cos(\lambda_\kappa \tau) \ket{\theta_1,\varphi_1 + \omega_\kappa \tau} \otimes \ket{\theta_2,\varphi_2 + \omega_\kappa \tau} \otimes \ket{\theta_3,\varphi_3 + \omega_\kappa \tau} \\
    {} - i\sin(\lambda_\kappa \tau) \ket{-\theta_1,\varphi_1 + \omega_\kappa \tau} \otimes \ket{-\theta_2,\varphi_2 + \omega_\kappa \tau} \otimes \ket{-\theta_3,\varphi_3 + \omega_\kappa \tau}.
\end{multline}
\end{linenomath}

From the above formula, we can see that the existence of the fundamental limit
for the precision with which we can measure time will cause the system initially
in a non-entangled state to evolve in time into an entangled state.

\section{Conclusions and Discussion}
\label{sec:5}
We developed a theory of non-relativistic quantum mechanics exhibiting a
non-zero minimal uncertainty in time. The theory was based on the Page--Wootters
formalism with a modified commutation relations between the time and frequency
operators. We received a modified version of the Schr\"odinger equation in a
continuous and discrete time representation. Interestingly, both representations
are equivalent, so, in particular, by solving the discrete Schr\"odinger
equation, we receive a time evolution on a lattice of a state vector, which we
can then transform to the continuous time representation. It should be noted
that even though we have a discrete time evolution, it does not mean that
fundamentally time is discrete. It only means that, using a particular
representation, we can describe time evolution on a lattice.

The developed formalism was used to describe time evolution of couple simple
quantum systems. The received results could be tested experimentally, at least
in principle. In particular, we found that in the system of spins-$1/2$ in an
external magnetic field, the frequency of precession of the spins depends on
the number of spins composing the system. Furthermore, the existence of a
fundamental limit with which we can measure time causes the entanglement of the
system of spins. These results are quite peculiar, especially that the spins do
not need to explicitly interact with each other, but still there exists some
kind of influence between them. The interpretation of these results can be
difficult and may suggest that perhaps the considered modification of
commutation relations is not physically correct and some different modification
will be more suitable.

It should be noted that we have not explicitly assumed that the parameter
$\kappa$ describing the minimal time scale is dependent on the system undergoing
time evolution. However, in view of the comment from the previous paragraph it
might be possible that $\kappa$ will depend on the system. Such a possibility
will require further investigation.

In this paper, we only considered time-independent Hamiltonians. It is possible
to generalize the theory to include time-dependent Hamiltonians by considering a
constraint operator $\hat{J}$ as in \eqref{eq:33}. However, in this case, it
will not be possible to use Theorem~\ref{thm:1} when deriving time evolution
equations because operators $\hbar\hat{\Omega} \otimes \hat{1}_S$ and
$\hat{H}_S(\hat{T})$ composing the constraint operator $\hat{J}$ do not commute.

Worth investigating will also be a relativistic quantum mechanics exhibiting a
non-zero uncertainty in time and space. In the Page--Wootters formalism, time is
treated as a quantum degree of freedom resulting from a correlation between the
``clock'' and the rest of the system. Interestingly the same approach can be
performed to describe space degrees of freedom \cite{Singh:2022}. Such formalism
will have an advantage of treating time and space on an equal footing. Lorentz
transformations can be defined as an appropriate unitary transformation. It is
then possible to consider different modifications, commonly found in the
literature, of the commutation relations between $\hat{T}$ and $\hat{\Omega}$
operators, as well as position and momentum operators.

\vspace{+6pt}

\funding{This research was funded by the Pozna\'n University of Technology under
Grant no. 0213/SBAD/0119.}

\dataavailability{The original contributions presented in this study are included
in the article; further inquiries can be directed to the author.}

\conflictsofinterest{The author declares no conflicts of interest.}

\appendixtitles{no}
\appendixstart
\appendix
\section[\appendixname~\thesection]{}\label{sec:AA}
In the following section, we will prove the technical theorem used in
Section~\ref{subsec:3.3}.

\begin{Theorem}
\label{thm:1}
Let $\hat{A}$ and $\hat{B}$ be (possibly unbounded) self-adjoint operators on
the Hilbert space $\mathcal{H}$ commuting with each other and let
$f \colon \sigma(\hat{A}) \cup \sigma(\hat{B}) \to \mathbb{C}$ be a bounded
continuous function defined on the joint spectrum of the operators $\hat{A}$
and $\hat{B}$. If for some vector $\ket{\psi} \in \mathcal{H}$,
\begin{equation}
\hat{A}\ket{\psi} = \hat{B}\ket{\psi},
\end{equation}
then
\begin{equation}
f(\hat{A})\ket{\psi} = f(\hat{B})\ket{\psi}.
\end{equation}
\end{Theorem}

The proof of the above theorem will require the following lemma.

\begin{Lemma}
\label{lem:1}
Let $\hat{A}$ and $\hat{B}$ be bounded operators on the Hilbert space
$\mathcal{H}$, and $\hat{A}^\dagger$ and $\hat{B}^\dagger$ their adjoints.
Let us assume that all those four operators pair-wise commute with each other.
Let $g \colon \sigma(\hat{A}) \cup \sigma(\hat{B}) \to \mathbb{C}$ be a
continuous function defined on the joint spectrum of the operators $\hat{A}$
and $\hat{B}$. If for some vector $\ket{\psi} \in \mathcal{H}$,
\begin{equation}
\hat{A}\ket{\psi} = \hat{B}\ket{\psi} \text{ and } \hat{A}^\dagger\ket{\psi} = \hat{B}^\dagger\ket{\psi},
\end{equation}
then
\begin{equation}
g(\hat{A})\ket{\psi} = g(\hat{B})\ket{\psi}.
\end{equation}
\end{Lemma}

\begin{proof}
The lemma is easily seen to be true for any complex monomial and by linearity
for any complex polynomial. For example, let $g(z) = z^n \bar{z}^m$; then,
\begin{linenomath}
\begin{align}
g(\hat{A})\ket{\psi} & = \hat{A}^n (\hat{A}^\dagger)^m \ket{\psi}
= \hat{A}^n (\hat{A}^\dagger)^{m-1} \hat{B}^\dagger \ket{\psi}
= \hat{B}^\dagger \hat{A}^n (\hat{A}^\dagger)^{m-1} \ket{\psi}  \nonumber \\
& = \hat{B}^\dagger \hat{A}^n (\hat{A}^\dagger)^{m-2} \hat{B}^\dagger \ket{\psi}
= (\hat{B}^\dagger)^2 \hat{A}^n (\hat{A}^\dagger)^{m-2} \ket{\psi}
= \dotsb = (\hat{B}^\dagger)^m \hat{A}^n \ket{\psi}  \nonumber \\
& = (\hat{B}^\dagger)^m \hat{A}^{n-1} \hat{B} \ket{\psi}
= \hat{B} (\hat{B}^\dagger)^m \hat{A}^{n-1} \ket{\psi}
= \hat{B} (\hat{B}^\dagger)^m \hat{A}^{n-2} \hat{B} \ket{\psi} \nonumber \\
& = \hat{B}^2 (\hat{B}^\dagger)^m \hat{A}^{n-2} \ket{\psi}
= \dotsb = \hat{B}^n (\hat{B}^\dagger)^m \ket{\psi}
= g(\hat{B})\ket{\psi}.
\end{align}
\end{linenomath}

Let now $g$ be a general continuous function.
Since $\hat{A}$ and $\hat{B}$ are bounded operators, their spectra
$\sigma(\hat{A})$ and $\sigma(\hat{B})$ are compact subsets of $\mathbb{C}$.
From the Stone--Weierstrass theorem, any continuous function defined on a
compact subset of $\mathbb{C}$ can be uniformly approximated by complex
polynomials. Therefore, the function $g$ is a uniform limit of a sequence of
complex polynomials $g_n$. The operators $\hat{A}$ and $\hat{B}$ are normal
since they commute with their adjoints. Using the continuous functional calculus
for normal bounded operators, we can define $g(\hat{A})$ and $g(\hat{B})$, and
the sequences of operators $g_n(\hat{A})$ and $g_n(\hat{B})$ converge uniformly
(in the operator norm) to $g(\hat{A})$ and $g(\hat{B})$, respectively. Hence,
\begin{equation}
g(\hat{A})\ket{\psi} = \lim_{n \to \infty} g_n(\hat{A})\ket{\psi}
= \lim_{n \to \infty} g_n(\hat{B})\ket{\psi} = g(\hat{B})\ket{\psi},
\end{equation}
which proves the lemma.
\end{proof}

\begin{proof}[Proof of Theorem~\ref{thm:1}]
The operators $\hat{A}$ and $\hat{B}$ are self-adjoint; therefore, their spectra
are subsets of $\mathbb{R}$. Hence, the following operators:
\begin{equation}
(\hat{A} - i\hat{1})^{-1}, \quad (\hat{A} + i\hat{1})^{-1}, \quad
(\hat{B} - i\hat{1})^{-1}, \quad (\hat{B} + i\hat{1})^{-1}
\label{eq:31}
\end{equation}
are resolvents of $\hat{A}$ and $\hat{B}$ and as such they are bounded
operators. Since $\hat{A}$ and $\hat{B}$ commute, all of the above operators
pair-wise commute with each other. Moreover,
\begin{equation}
\bigl((\hat{A} - i\hat{1})^{-1}\bigr)^\dagger = (\hat{A} + i\hat{1})^{-1} \text{ and }
\bigl((\hat{B} - i\hat{1})^{-1}\bigr)^\dagger = (\hat{B} + i\hat{1})^{-1}.
\end{equation}

Using the commutativity of $\hat{A}$ and $\hat{B}$, we also calculate that
\begin{linenomath}
\begin{gather}
\hat{B}\ket{\psi} = \hat{A}\ket{\psi} \quad\iff\quad
(\hat{B} - i\hat{1})\ket{\psi} = (\hat{A} - i\hat{1})\ket{\psi} \quad\iff\quad \nonumber \\
\ket{\psi} = (\hat{B} - i\hat{1})^{-1} (\hat{A} - i\hat{1})\ket{\psi} \quad\iff\quad
\ket{\psi} = (\hat{A} - i\hat{1}) (\hat{B} - i\hat{1})^{-1} \ket{\psi} \quad\iff\quad \nonumber \\
(\hat{A} - i\hat{1})^{-1} \ket{\psi} = (\hat{B} - i\hat{1})^{-1} \ket{\psi}
\end{gather}
\end{linenomath}
and similarly
\begin{equation}
(\hat{A} + i\hat{1})^{-1} \ket{\psi} = (\hat{B} + i\hat{1})^{-1} \ket{\psi}.
\end{equation}

Therefore, all four operators \eqref{eq:31} satisfy conditions of
Lemma~\ref{lem:1}.

Let $g(z) = f(1/z + i)$ for $z \neq 0$ and $g(0) = 0$ be a function defined on
a joint spectrum of the resolvents $(\hat{A} - i\hat{1})^{-1}$ and
$(\hat{B} - i\hat{1})^{-1}$. This joint spectrum is a subset of a compact set
\begin{equation}
X = \{z \in \mathbb{C} \mid z = 1/(x - i) \text{ for all $x \in \mathbb{R}$}\} \cup \{0\}.
\end{equation}

The set $X$ is just a circle of radius $1/2$ centered at $i/2$. The function $g$
is bounded and continuous everywhere except possibly at point $z = 0$. To make
this function continuous, let $h_n$ be a compactly supported continuous function
defined on $\mathbb{R}$ such that $0 \leq h_n(x) \leq 1$ for all
$x \in \mathbb{R}$ and $h_n(x) = 1$ for $x \in [-n,n]$. Then, the functions
$g_n(z) = h_n(1/z + i)g(z)$ for $n \in \mathbb{N}$ are everywhere continuous,
$\lim_{n \to \infty} g_n(z) = g(z)$ pointwise, and $\norm{g_n}_\infty \leq
\norm{g}_\infty$. Using the Borel functional calculus for normal bounded
operators, we can define $g((\hat{A} - i\hat{1})^{-1})$ and
$g((\hat{B} - i\hat{1})^{-1})$, and the sequences of operators
$g_n((\hat{A} - i\hat{1})^{-1})$ and $g_n((\hat{B} - i\hat{1})^{-1})$ converge
strongly to $g((\hat{A} - i\hat{1})^{-1})$ and $g((\hat{B} - i\hat{1})^{-1})$
respectively. Hence, with the help of Lemma~\ref{lem:1}
\begin{linenomath}
\begin{align}
g\bigl((\hat{A} - i\hat{1})^{-1}\bigr)\ket{\psi} & =
    \lim_{n \to \infty} g_n\bigl((\hat{A} - i\hat{1})^{-1}\bigr)\ket{\psi}
= \lim_{n \to \infty} g_n\bigl((\hat{B} - i\hat{1})^{-1}\bigr)\ket{\psi} \nonumber \\
& = g\bigl((\hat{B} - i\hat{1})^{-1}\bigr)\ket{\psi}.
\end{align}
\end{linenomath}

From this follows
\begin{equation}
f(\hat{A})\ket{\psi} = f(\hat{B})\ket{\psi},
\end{equation}
which proves the theorem.
\end{proof}

%%%%%%%%%%%%%%%%%%%%%%%%%%%%%%%%%%%%%%%%%%%%%%%%%%%%%%%%%%%%%%%%%%%%%%%%%%%%%%%%
\begin{adjustwidth}{-\extralength}{0cm}
\reftitle{References}

%\PublishersNote{}
\end{adjustwidth}
\end{document}